\documentclass[%
reprint,
nofootinbib,
 amsmath,amssymb,
 aps,
floatfix,
]{revtex4-2}

\usepackage{graphicx}
\usepackage{dcolumn}
\usepackage{bm}
\usepackage{float} %
\usepackage{placeins}
\usepackage{url}

\begin{document}

\preprint{arXiv:2601.19616}
\preprint{Published in Phys.\ Rev.\ D; \url{https://doi.org/10.1103/f3jn-7wcp}}

\title{Rarity of rocket-driven Penrose extraction in Kerr spacetime}
\thanks{Corresponding author: An T. Le, an.lt@vinuni.edu.vn; an@robot-learning.de}%

\author{An T. Le}
\homepage{https://vinuni.edu.vn/people/le-thai-an/}
\affiliation{
 College of Engineering and Computer Sciences, VinUniversity, Hanoi, Vietnam
}%
\affiliation{
 Center for Environmental Intelligence, VinUniversity, Hanoi, Vietnam
}%
\affiliation{
 Intelligent Autonomous Systems, TU Darmstadt, Germany
}%

\begin{abstract}
We study rocket-driven Penrose extraction in the test-particle limit on a fixed Kerr background for equatorial prograde flybys under explicit steering prescriptions. A spacecraft ejects exhaust inside the ergosphere; when the exhaust attains negative Killing energy, the remaining spacecraft gains energy by 4-momentum conservation. Across 320{,}000 simulated trajectories spanning black-hole spin, exhaust velocity, and orbital parameters, extraction with escape is rare in broad parameter scans (at most ${\sim}1\%$) and requires high spin ($a/M\gtrsim 0.89$), highly relativistic exhaust ($v_e\gtrsim 0.91c$), and finely tuned initial conditions. Under optimal tuning the success rate reaches ${\sim}70\%$ at $a/M = 0.95$. For representative escape trajectories, a single periapsis impulse is more propellant-efficient than the continuous-thrust controllers studied here. All quoted thresholds are empirical and specific to the orbit family, prior, and steering protocol studied.
\end{abstract}

\keywords{Penrose process, Kerr black holes, energy extraction, relativistic propulsion, Monte Carlo simulation}
\maketitle


\section{Introduction} \label{sec:intro}

The extraction of energy from rotating black holes stands as one of the most remarkable predictions of general relativity. In 1971, Penrose and Floyd described a mechanism whereby particles entering the ergosphere of a Kerr black hole could split, with one fragment falling into the horizon carrying negative Killing energy while the other escapes with energy exceeding that of the original particle~\cite{penrose1971extraction}. The rotational energy of a Kerr black hole above its irreducible mass amounts to up to 29\% of the total mass~\cite{christodoulou1970reversible, bardeen1972rotating}, in principle accessible through repeated Penrose events; the maximum single-decay efficiency is $\sim$20.7\% for extremal Kerr~\cite{wald1974energy}.

The theoretical elegance of the Penrose process has inspired decades of research into its variants and astrophysical manifestations. Piran, Shaham, and Katz~\cite{piran1975high} showed that particle \emph{collisions} in the ergosphere could achieve substantially higher center-of-mass energies. This collisional Penrose process was later shown by Ba\~{n}ados, Silk, and West~\cite{banados2009kerr} to yield arbitrarily high center-of-mass energies near extremal horizons, though practical efficiencies are limited by gravitational redshift of escaping debris~\cite{bejger2012collisional, berti2015ultrahigh, schnittman2014revised}. Leiderschneider and Piran~\cite{leiderschneider2016maximal} systematically characterized the maximal efficiency of the collisional process.

In astrophysical contexts, electromagnetic variants of energy extraction have proven more observationally relevant. The Blandford--Znajek mechanism~\cite{blandford1977electromagnetic} extracts rotational energy through horizon-threading magnetic fields supported by surrounding accretion flows, powering the relativistic jets observed in active galactic nuclei and quasars~\cite{tchekhovskoy2011efficient}. More recently, Comisso and Asenjo~\cite{comisso2021magnetic} demonstrated that magnetic reconnection in the ergosphere provides an efficient Penrose-like mechanism, potentially explaining the rapid variability of high-energy emission from supermassive black holes. Koide and Arai~\cite{koide2008energy} had earlier explored similar magnetic reconnection scenarios numerically.

Despite this extensive body of work, a fundamental question remains largely unexplored: \emph{How fine-tuned must conditions be for successful Penrose extraction with escape to infinity?} The classical Penrose process requires precisely tuned initial conditions---the decaying particle must enter the ergosphere on a trajectory that (i) penetrates deep enough for negative-energy states to be accessible (the captured fragment must carry negative energy where frame-dragging is strong), (ii) does not plunge into the horizon, and (iii) allows the positive-energy fragment to escape with sufficient angular momentum. This narrow viable region in parameter space has been noted qualitatively~\cite{chandrasekhar1985mathematical, bhat1985energetics}, but to our knowledge no systematic statistical characterization exists to quantify \emph{how rare} success is or \emph{what parameter thresholds} govern extraction.

This paper addresses this gap by studying the Penrose process via \emph{rocket propulsion}, in which a spacecraft ejects mass (exhaust) inside the ergosphere, analogous to the classical particle decay but with thrust direction and magnitude allowed to vary under explicit steering prescriptions. This is a kinematic test-particle study: the spacecraft mass satisfies $m_0 \ll M$, so the Kerr background is fixed and self-force effects are negligible; inferred changes to the hole's rotational energy are post-hoc estimates, not dynamical backreaction. We restrict attention to \emph{initially unbound equatorial prograde flybys}, the natural baseline family for the mission objective of energy gain with escape to infinity: these trajectories enter the ergosphere, interact with the black hole's rotation, and, if conditions are favorable, escape to infinity with a net energy increase. (Later diagnostic sweeps in Sec.~\ref{sec:ensemble} also explore a broader $(E_0, L_z)$ domain that intentionally includes bound and retrograde states to map failure boundaries; those extended-domain results characterize the parameter landscape, not the targeted mission family.) We provide a \emph{computational feasibility map} for rocket-like Penrose extraction, establishing quantitative thresholds, efficiencies, and success probabilities across large parameter sweeps. Our formulation offers three advantages. First, it connects the abstract Penrose process to a concrete engineering scenario, clarifying which physical parameters govern extraction. Second, it enables Monte Carlo exploration of the initial condition space, quantifying the statistical rarity of successful extraction with escape. Third, it permits comparison of thrust strategies (impulsive vs.\ continuous), revealing path-averaging effects that reduce efficiency in extended maneuvers. The novelty of this work lies not in the existence of the Penrose process, but in providing \emph{quantified success rates and parameter thresholds} under a rocket-propulsion control model, together with a thrust-profile comparison and a constrained optimality argument. Our primary contribution is the impulsive periapsis-ensemble rarity and threshold map; the continuous-thrust comparison is a secondary study of representative matched trajectories under specific implemented controllers.

Our central finding is a set of \emph{constraint results}: within our equatorial prograde-flyby family and steering prescriptions (hereafter our \emph{baseline family and protocol}), successful Penrose extraction with spacecraft escape is observed only for (i) high spin ($a/M \gtrsim 0.89$), (ii) highly relativistic exhaust ($v_e \gtrsim 0.91c$), and (iii) carefully tuned initial conditions within a narrow ``sweet spot.'' Under maximum tuning the success rate can reach $\sim$70\% at $a/M = 0.95$, but this represents the fine-tuned limit, not typical behavior.\footnote{An exploratory scan at $a/M = 0.99$ yields ${\sim}65$--$70\%$ under the same prior (independent-seed verification; see Appendix~\ref{app:exploratory-99}).} Single periapsis-impulse ejection is more propellant-efficient than continuous thrust for the representative matched escape trajectories studied, consistent with the path-averaging penalty of the fixed-worldline optimality argument and, additionally, with trajectory-deformation effects during extended ergosphere transit. Quantitative details, confidence intervals, and efficiency metrics are given in the results sections.

Throughout this paper we distinguish three logically separate layers: (i) \emph{local kinematics}, whether the exhaust attains negative Killing energy ($E_{\rm ex} < 0$); (ii) \emph{mission outcome}, whether the spacecraft both gains energy and escapes to infinity ($\mathcal{S} = (\Delta E > 0) \land \text{escape}$, Eq.~\ref{eq:penrose-success}); and (iii) \emph{performance metrics}, cumulative and traditional efficiencies ($\eta_{\rm cum}$, $\eta_{\rm trad}$). These are defined in Sec.~\ref{sec:continuous-penrose}\,D--E and applied throughout the results.

The paper is organized as follows. Section~\ref{sec:problem} reviews Kerr spacetime geometry. Section~\ref{sec:classical-penrose} summarizes the classical Penrose process, and Section~\ref{sec:continuous-penrose} introduces the rocket-driven variant. Sections~\ref{sec:formulation}--\ref{sec:thrust-model} present the mathematical formulation and thrust model. Section~\ref{sec:numerical-methods} describes numerical methods. Section~\ref{sec:validation} presents numerical results, including a theoretical optimality argument for single-impulse extraction (Sec.~\ref{sec:optimality}) followed by an illustrative controller comparison. The statistical core of this study is the periapsis-impulse ensemble (Tables~\ref{tab:broad-outcomes}--\ref{tab:thrust-params}; exploratory $a/M=0.99$ results appear in Appendix~\ref{app:exploratory-99}); continuous thrust is examined as a controller comparison on representative matched trajectories (Sec.~\ref{sec:thrust-comparison}), not as a coequal experimental pillar. Section~\ref{sec:conclusions} summarizes implications.

\section{Problem Setting: Kerr Spacetime}
\label{sec:problem}

To make the control and energetics statements precise, we begin by fixing notation for Kerr geometry, the ergosphere, and the conserved quantities used throughout.

We consider a rotating (Kerr) black hole of mass $M$ and angular momentum $J = aM$, where $0 \leq a \leq M$ is the spin parameter. In Boyer--Lindquist coordinates $(t, r, \theta, \phi)$, the Kerr metric takes the form
\begin{align}
\label{eq:kerr-metric}
ds^2 = &-\left(1 - \frac{2Mr}{\Sigma}\right)dt^2 - \frac{4Mar\sin^2\theta}{\Sigma}dt\,d\phi \;+ \nonumber \\
&\frac{\Sigma}{\Delta}dr^2 + \Sigma\,d\theta^2 + \frac{A\sin^2\theta}{\Sigma}d\phi^2,
\end{align}
where
\begin{align}
\Sigma &= r^2 + a^2\cos^2\theta, \\
\Delta &= r^2 - 2Mr + a^2, \\
A &= (r^2 + a^2)^2 - a^2\Delta\sin^2\theta.
\end{align}

The event horizon is located at
\begin{equation}
r_+ = M + \sqrt{M^2 - a^2},
\end{equation}
which exists for $a \leq M$. The extremal limit $a = M$ corresponds to a maximally rotating black hole with $r_+ = M$.

\subsection{Constants of Motion}

The Kerr spacetime admits two Killing vectors: $\xi^\mu = (\partial/\partial t)^\mu$ (stationarity) and $\psi^\mu = (\partial/\partial \phi)^\mu$ (axisymmetry). For geodesic motion, these yield conserved quantities:
\begin{align}
E &= -p_t = -g_{t\mu}p^\mu, \label{eq:energy-def} \\
L_z &= p_\phi = g_{\phi\mu}p^\mu, \label{eq:Lz-def}
\end{align}
where $E$ is the Killing energy (as measured by observers at rest at infinity) and $L_z$ is the azimuthal angular momentum. We normalize the spacecraft's initial rest mass to $m_0 = 1$ throughout (see Sec.~\ref{sec:numerical-methods}); all sweep parameters $E_0 \equiv E/m_0 = -u_t$ and $L_{z,0} \equiv L_z/m_0 = u_\phi$ denote \emph{specific} (per-unit-mass) quantities. In particular, an unbound orbit has specific energy $E_0 > 1$ (not merely $E_0 > 0$), and escape to infinity requires $E/m > 1$ at large radius. Throughout this paper, ``retrograde'' refers to negative $L_z$ (opposing the black hole rotation), not to $u^\phi < 0$ in Boyer--Lindquist coordinates; inside the ergosphere, frame dragging forces all future-directed worldlines to have $u^\phi > 0$.

\subsection{The Ergosphere}

The stationary limit surface (outer boundary of the ergosphere) is defined by $g_{tt} = 0$, yielding
\begin{equation}
\label{eq:ergosphere}
r_{\rm erg}(\theta) = M + \sqrt{M^2 - a^2\cos^2\theta}.
\end{equation}
At the equatorial plane ($\theta = \pi/2$), we have $r_{\rm erg} = 2M$ for all values of spin. The ergosphere is the region $r_+ < r < r_{\rm erg}(\theta)$ where $\xi^\mu$ becomes spacelike, meaning no observer can remain stationary. Inside the ergosphere, timelike worldlines can have $E < 0$ (negative Killing energy); outside, the timelike character of $\xi^\mu$ forces $E \ge 0$ for all physical worldlines, so negative-energy states are confined to the ergoregion. Detailed properties of negative-energy geodesics in the Kerr ergosphere are discussed in Ref.~\cite{grib2014negative}.

The angular velocity of frame dragging is
\begin{equation}
\omega = -\frac{g_{t\phi}}{g_{\phi\phi}} = \frac{2Mar}{A}.
\end{equation}
This frame-dragging is central to energy extraction via the Penrose mechanism.

\section{Classical Penrose Process}
\label{sec:classical-penrose}

The original Penrose process~\cite{penrose1971extraction} provides a mechanism to extract rotational energy from a Kerr black hole using particle decay.

\subsection{Physical Mechanism}

Consider an incident particle with 4-momentum $p^\mu_0$ and energy $E_0 > 0$ entering the ergosphere. Inside the ergosphere, the particle undergoes a decay:
\begin{equation}
\text{particle}_0 \rightarrow \text{particle}_1 + \text{particle}_2.
\end{equation}
Conservation of 4-momentum requires
\begin{equation}
p^\mu_0 = p^\mu_1 + p^\mu_2.
\end{equation}
Contracting with the Killing vector $\xi_\mu = (\partial/\partial t)_\mu$:
\begin{equation}
E_0 = E_1 + E_2.
\end{equation}

Inside the ergosphere, timelike worldlines can have $E < 0$ (negative Killing energy, as measured by observers at rest at infinity) while remaining physical. Specifically, if particle~2 is ejected with large retrograde angular momentum (opposing the black hole rotation), it can achieve $E_2 < 0$. Particle~1 then escapes with
\begin{equation}
E_1 = E_0 - E_2 > E_0,
\end{equation}
extracting energy from the black hole.

\subsection{Conditions for Energy Extraction}

For successful energy extraction, the escaping particle must satisfy:
\begin{enumerate}
\item The decay occurs inside the ergosphere: $r_+ < r < r_{\rm erg}$.
\item The captured particle has negative Killing energy: $E_2 < 0$.
\item The escaping fragment follows a geodesic that reaches spatial infinity; for a massive particle, this requires specific energy $E_1/m_1 > 1$ (unbound) together with an allowed radial turning-point structure. Positive total energy $E_1 > 0$ alone is insufficient, since positive-energy bound orbits exist in Kerr.
\end{enumerate}

A necessary condition for $E_2 < 0$ is that the fragment is created inside the ergosphere and carries sufficiently negative (retrograde) angular momentum. Writing $p^\mu = m u^\mu$ and $u^\phi = \Omega\,u^t$ with $\Omega := d\phi/dt$, the condition $E<0$ requires
\begin{equation}
g_{tt}+g_{t\phi}\Omega>0
\qquad\Longleftrightarrow\qquad
\Omega < -\frac{g_{tt}}{g_{t\phi}} \quad (g_{t\phi}<0).
\end{equation}
Even inside the ergosphere, frame dragging enforces $\Omega\in(\Omega_-,\Omega_+)$ with $\Omega_\pm>0$, where
\begin{equation}
\Omega_\pm = \frac{-g_{t\phi} \pm \sqrt{g_{t\phi}^2 - g_{tt}g_{\phi\phi}}}{g_{\phi\phi}}.
\end{equation}

\subsection{Efficiency Limits}

Two distinct efficiency bounds govern energy extraction. The \emph{total extractable rotational energy} of a Kerr black hole is~\cite{christodoulou1970reversible}:
\begin{equation}
\label{eq:rotational-bound}
\eta_{\rm rot} = 1 - \sqrt{\frac{1 + \sqrt{1 - (a/M)^2}}{2}},
\end{equation}
which gives $\eta_{\rm rot} \approx 29.3\%$ for an extremal black hole ($a = M$). This represents the thermodynamic limit on cumulative extraction via repeated processes.

The \emph{single-decay efficiency} was bounded by Wald~\cite{wald1974energy}. For an incident particle from rest at infinity in extremal Kerr ($a = M$), the maximum fractional energy gain of the escaping fragment is
\begin{equation}
\label{eq:wald-limit}
\eta_{\rm Wald} = \frac{E_1 - E_0}{E_0} \approx 20.7\%.
\end{equation}
Our single-impulse results should be compared to this classical benchmark, while cumulative processes may approach the rotational bound through multiple extractions.

\section{Rocket-Driven Penrose Variant}
\label{sec:continuous-penrose}

We formulate a general rocket-driven Penrose model with two thrust modes: single periapsis impulses and continuous thrust. Most ensemble statistics reported below use the impulsive mode; continuous thrust is used mainly for matched-trajectory comparisons (Sec.~\ref{sec:thrust-comparison}). The continuous formulation reveals quantitative path-averaging effects absent from single-decay analyses.

\subsection{Physical Concept}

A spacecraft with initial rest mass $m_0$ and specific energy $E_0 = -u_t > 1$ (unbound) enters the ergosphere on a \emph{prograde flyby orbit} (angular momentum $L_z > 0$, co-rotating with the black hole). The orbit is chosen such that its periapsis lies inside the ergosphere. While passing through the ergosphere near periapsis, the spacecraft activates a rocket engine that:
\begin{enumerate}
\item Ejects mass at exhaust velocity $v_e$ (in the spacecraft's rest frame).
\item Chooses a thrust orientation such that the exhaust carries retrograde angular momentum ($L_{z,\rm ex} < 0$).
\item Inside the ergosphere, the exhaust can then have $E_{\rm ex} < 0$, transferring energy to the remaining spacecraft.
\end{enumerate}

Conceptually, each infinitesimal exhaust element plays the role of the captured negative-energy fragment in the classical single-decay, while the remaining spacecraft plays the role of the escaping fragment; the rocket engine provides control over the split direction and timing.

In successful extraction-with-escape cases, the ejection of negative-energy exhaust results in the spacecraft gaining energy at infinity, emerging from the flyby with $E_f > E_0$. This is the rocket-driven Penrose process: many infinitesimal mass ejections, each of which \emph{can} carry away negative Killing energy when the exhaust direction is suitably chosen.

\textbf{Critical insight:} In our parameter sweeps, successful extraction with escape arises from \emph{prograde, initially unbound} flybys ($L_z>0$, $E_0>1$). Retrograde ($L_z<0$) or initially bound ($E_0<1$) initial conditions did not yield successful extraction with escape in our sampled domain, which may reflect their more restrictive turning-point structure for deep ergosphere transits.

\subsection{Advantages Over Classical Process}

Our continuous formulation offers several modeling advantages: rocket propulsion is a well-understood technology analogy; thrust magnitude and direction can be modulated; and active propulsion can be used to attempt to prolong ergosphere dwell time and reduce capture risk under explicit steering prescriptions.

\subsection{Definition of Extraction-with-Escape Success}

We define a \emph{successful extraction-with-escape event} (hereafter ``extraction-with-escape success,'' denoting the mission-level criterion, not local Penrose extraction alone) as the conjunction of two conditions:
\begin{align}
\label{eq:penrose-success}
\mathcal{S} := &(\Delta E > 0) \;\land\; (\text{spacecraft escapes to infinity}).
\end{align}
Operationally, as a finite-radius escape proxy, we diagnose \emph{escape} when the spacecraft reaches $r > 50M$ with outward radial velocity $dr/d\tau > 0$ and specific energy $E/m > 1$. We diagnose \emph{capture} when $r < r_+ + 0.02M$ (horizon approach) or when integration fails due to stiffness near the horizon. We conservatively count integration failures (11.2\% of samples) as captures.

Under our throttle and sign-selection policy, all single-impulse runs satisfying $\Delta E > 0$ with escape also exhibit $E_{\rm ex} < 0$ during the burn: within the sampled single-impulse ensemble, $E_{\rm ex} < 0$ is a \emph{necessary} condition for extraction-with-escape success. However, $E_{\rm ex} < 0$ is \emph{not sufficient} for mission success. The continuous-thrust capture example in Table~\ref{tab:thrust-comparison} achieves $E_{\rm ex} < 0$ for nearly all thrust events and $\Delta E > 0$, yet the spacecraft does not escape. More generally, negative-energy exhaust is confined to the ergoregion and is captured (Sec.~\ref{sec:problem}). The spacecraft must also navigate out of the ergosphere without plunging. We therefore use $\Delta E > 0 \,\land\, \text{escape}$ as the primary mission-level criterion, with $E_{\rm ex} < 0$ serving as the local Penrose signature.

\subsection{Efficiency Metrics}

We define the \emph{instantaneous efficiency} $\eta_{\rm inst} = dE/d(-m)$, where $\eta_{\rm inst} > 1$ indicates propellant rest-mass break-even is exceeded (the Killing-energy gain per unit rocket mass loss exceeds unity). However, this alone does not prove Penrose extraction; a direct signature is negative exhaust Killing energy $E_{\rm ex}<0$ inside the ergosphere.

The \emph{cumulative efficiency} over the maneuver is:
\begin{equation}
\label{eq:eta-cum}
\eta_{\rm cum} = \frac{E_f - E_0}{m_0 - m_f} = \frac{\Delta E}{\Delta m}.
\end{equation}

For later parameter sweeps we report propellant expenditure via the ejected rocket mass fraction
\begin{equation}
\delta m := \frac{m_0 - m_f}{m_0},
\end{equation}
so that $\Delta m = m_0 - m_f = \delta m\,m_0$.

For relativistic exhaust velocity $v_e$ with Lorentz factor $\gamma_e = 1/\sqrt{1-v_e^2}$, the \emph{differential} relation $-dm = \gamma_e\,d\mu$ between rocket mass loss and exhaust rest mass is exact in the instantaneous rest frame and applies exactly to continuous-thrust simulations. For a finite (impulsive) burn, however, the exact mass-shell norm of $p'_\mu = p_\mu - \delta\mu\,u_{{\rm ex},\mu}$ gives
\begin{equation}
\label{eq:finite-impulse-mass}
m_f^2 = m_0^2 - 2\gamma_e m_0\,\delta\mu + \delta\mu^2,
\end{equation}
so the exact exhaust rest mass for a desired mass fraction $\delta m = (m_0 - m_f)/m_0$ is
\begin{equation}
\label{eq:dmu-exact}
\delta\mu = m_0\!\left[\gamma_e - \sqrt{\gamma_e^2 - 2\delta m + \delta m^2}\right].
\end{equation}
Our single-impulse simulations parameterize by a \emph{nominal burn parameter} $\delta m_{\rm nom}$ (not a physical mass fraction; see below) and compute $\delta\mu = \delta m_{\rm nom}\cdot m_0/\gamma_e$ (the differential-limit approximation), then apply exact 4-momentum conservation (Eq.~\ref{eq:momentum-conservation}). Because this approximate $\delta\mu$ exceeds the exact value from Eq.~\eqref{eq:dmu-exact}, the actual rocket mass fraction consumed $\delta m_{\rm actual} = (m_0 - m_f)/m_0$ exceeds $\delta m_{\rm nom}$: at $v_e=0.95c$ and $\delta m_{\rm nom}=0.4$, the discrepancy reaches ${\sim}34\%$ (see Table~\ref{tab:thrust-comparison} caption for actual consumed fractions). The cumulative efficiency $\eta_{\rm cum} = \Delta E/(m_0 - m_f)$ is computed from the exact post-impulse mass and is unaffected by this labeling convention. For the per-exhaust-rest-mass ratio $\Delta E/\delta\mu$ reported in Table~\ref{tab:thrust-comparison}, we use the exact $\delta\mu$ from Eq.~\eqref{eq:dmu-exact} applied to the actual consumed fraction $\delta m_{\rm actual}$; this is the physically meaningful normalization. We also define the \emph{traditional Penrose efficiency}, analogous to the single-decay efficiency used for the classical process:
\begin{equation}
\label{eq:eta-trad}
\eta_{\rm trad} = \frac{E_f - E_0}{E_0},
\end{equation}
which measures the fractional energy gain relative to the initial energy. In Table~\ref{tab:thrust-comparison} we report $\eta_{\rm cum}$ and $\Delta E/\delta\mu$ for the representative trajectories; $\eta_{\rm trad}$ is retained as a complementary energy-normalized diagnostic and as the quantity directly comparable to Wald's classic single-decay benchmark. The cumulative efficiency $\eta_{\rm cum}$ (Eq.~\ref{eq:eta-cum}) is an \emph{engineering propellant-efficiency metric}: it quantifies how much Killing energy is gained per unit of rocket mass expended. It is the natural figure of merit for the rocket-propulsion context but mixes Penrose extraction physics with propellant expenditure. For cleaner Penrose-physics diagnostics, $E_{\rm ex}<0$ (the local kinematic signature of negative-energy exhaust) and $\eta_{\rm trad}$ (the fractional energy gain) are more appropriate; in particular, $\eta_{\rm trad}$ enables direct comparison with Wald's classic single-decay benchmark $\eta_{\rm Wald} \approx 20.7\%$ for extremal Kerr~\cite{wald1974energy}.

\subsection{Local Kinematic Threshold Estimate}
\label{sec:velocity-threshold}

A key prediction of our Monte Carlo study is a sharp velocity threshold near $v_e \approx 0.91$--$0.92c$, below which extraction-with-escape largely fails (local $E_{\rm ex}<0$ can still occur without the spacecraft escaping). Before formalizing the thrust model (Secs.~\ref{sec:formulation}--\ref{sec:thrust-model}), we preview the physical insight behind this threshold by connecting it to the local requirement that the \emph{exhaust} (not the spacecraft) attain negative Killing energy. The exact exhaust-energy expression derived in Sec.~\ref{sec:momentum-conservation} is:
\begin{equation}
E_{\rm ex} = -\gamma_e\!\left(u_t - v_e s_t\right),
\end{equation}
where $u_t = g_{t\mu}u^\mu$ is the covariant time component of the spacecraft 4-velocity and $s_t = g_{t\mu}s^\mu$ is the covariant time component of the unit spacelike \emph{thrust} direction (the exhaust is emitted opposite to $s^\mu$).

Negative-energy exhaust requires
\begin{equation}
E_{\rm ex} < 0 \quad \Longleftrightarrow \quad u_t - v_e s_t > 0.
\label{eq:exhaust-neg-energy-condition}
\end{equation}
For the unbound, escaping trajectories of interest we have $E/m = -u_t > 1$, hence $u_t < 0$. Equation~\eqref{eq:exhaust-neg-energy-condition} then implies a simple but important sign constraint: the thrust direction must satisfy $s_t < 0$ so that the $-v_e s_t$ term can overcome the negative $u_t$. In our setup, this is achieved by selecting thrust orientations that eject exhaust with sufficiently retrograde angular momentum (cf.\ the convention defined in Sec.~\ref{sec:problem}).

When $u_t<0$ and $s_t<0$, Eq.~\eqref{eq:exhaust-neg-energy-condition} becomes a \emph{lower bound} on exhaust speed:
\begin{equation}
v_e \;>\; v_e^{\rm crit} \;:=\; \frac{|u_t|}{|s_t|},
\label{eq:ve-crit}
\end{equation}
with $u_t$ and $s_t$ evaluated at the burn event. We stress that $v_e > v_e^{\rm crit}$ is a \emph{necessary local condition} for $E_{\rm ex} < 0$ at a given burn event, not a sufficient condition for mission-level extraction-with-escape success ($\mathcal{S}$, Eq.~\ref{eq:penrose-success}): the spacecraft must still escape to infinity with $\Delta E > 0$. This makes the origin of the sharp onset transparent: as $v_e$ increases, the set of spacetime points and thrust angles satisfying $v_e>v_e^{\rm crit}$ (hence $E_{\rm ex}<0$) expands rapidly.

Evaluating Eq.~\eqref{eq:ve-crit} at periapsis for representative sweet-spot states ($a/M=0.95$, $r_p \approx 1.5M$, $E_0 \approx 1.2$, $L_z \approx 3.0$) yields $v_e^{\rm crit} \approx 0.85$--$0.92c$, depending on the orbital phase and steering angle. This estimate is consistent with the Monte Carlo observation (Fig.~\ref{fig:thrust-sensitivity}) that success rates transition sharply around $v_e \approx 0.91$--$0.92c$: below this value, only rare geometries achieve $E_{\rm ex}<0$ while retaining escape; above it, negative-energy exhaust becomes robust over a wide range of periapsis configurations.

\textbf{Important caveat:} The threshold $v_e \approx 0.91$--$0.92c$ is \emph{not} a universal Kerr-invariant constant. It is an empirical threshold specific to our baseline Gaussian sweet-spot prior (Gaussian sampling around $E_0 = 1.22$, $L_z = 3.05$), our greedy thrust policy (maximize instantaneous $-dp_t/d\tau$), and the finite angular discretization of the thrust-direction search (74 candidate directions; Sec.~\ref{sec:protocol-summary}). Different orbit families, steering laws, spin parameters, or finer direction scans will yield different critical velocities. The threshold is thus a geometric constraint arising from the Killing-energy decomposition $v_e > |u_t|/|s_t|$ (Eq.~\ref{eq:ve-crit}), evaluated for a particular class of trajectories and controller.

We now formalize this controlled-ejection Penrose variant as a variable-mass, forced trajectory problem in Kerr spacetime.

\section{Mathematical Formulation}
\label{sec:formulation}

We now present the mathematical formulation. Geodesic (thrust-off) phases follow Hamilton's equations in Kerr spacetime; thrust-on phases are governed by a forced variable-mass equation of motion (Sec.~\ref{sec:thrust-model}).

\subsection{Equatorial Restriction}

For simplicity, we restrict to equatorial motion ($\theta = \pi/2$, $p_\theta = 0$). The dynamics reduce from 8 to 6 dimensions. At the equator:
\begin{align}
\Sigma &= r^2, \\
\Delta &= r^2 - 2Mr + a^2, \\
g_{tt} &= -\left(1 - \frac{2M}{r}\right), \\
g_{t\phi} &= -\frac{2Ma}{r}, \\
g_{rr} &= \frac{r^2}{\Delta}, \\
g_{\phi\phi} &= r^2 + a^2 + \frac{2Ma^2}{r}.
\end{align}

\subsection{State Variables}

The state vector for the spacecraft is
\begin{equation}
\bm{y} = (r, \phi, p_t, p_r, p_\phi, m),
\end{equation}
where:
\begin{itemize}
\item $r$: Boyer--Lindquist radial coordinate
\item $\phi$: Azimuthal angle
\item $p_t$: Covariant time momentum (energy: $E = -p_t$)
\item $p_r$: Covariant radial momentum
\item $p_\phi$: Covariant azimuthal momentum (angular momentum: $L_z = p_\phi$)
\item $m$: Rest mass (evolving due to propellant consumption)
\end{itemize}

\subsection{Mass-Shell Constraint}

The spacecraft 4-momentum must satisfy the mass-shell condition:
\begin{equation}
\label{eq:mass-shell}
g^{\mu\nu}p_\mu p_\nu = -m^2.
\end{equation}
Expanding at the equator:
\begin{equation}
g^{tt}p_t^2 + 2g^{t\phi}p_t p_\phi + g^{rr}p_r^2 + g^{\phi\phi}p_\phi^2 = -m^2.
\end{equation}
This constrains the relationship between $p_t$, $p_r$, and $p_\phi$ for given $m$.

\subsection{Hamiltonian}

The Hamiltonian for a free particle is
\begin{equation}
\mathcal{H} = \frac{1}{2}g^{\mu\nu}p_\mu p_\nu = \frac{1}{2}\left(g^{tt}p_t^2 + 2g^{t\phi}p_t p_\phi + g^{rr}p_r^2 + g^{\phi\phi}p_\phi^2\right).
\end{equation}
For massive particles with $p_\mu = m u_\mu$, the mass-shell constraint gives $\mathcal{H} = -m^2/2$.

For evolution in proper time $\tau$ (rather than affine parameter), we define:
\begin{equation}
\mathcal{H}_\tau = \frac{1}{2m}\left(g^{\mu\nu}p_\mu p_\nu + m^2\right),
\end{equation}
which satisfies $\mathcal{H}_\tau = 0$ on shell. Hamilton's equations in proper time are:
\begin{align}
\frac{dr}{d\tau} &= \frac{1}{m}\frac{\partial\mathcal{H}}{\partial p_r} = \frac{g^{rr}p_r}{m}, \label{eq:dr-dtau} \\
\frac{d\phi}{d\tau} &= \frac{1}{m}\frac{\partial\mathcal{H}}{\partial p_\phi} = \frac{g^{\phi\phi}p_\phi + g^{t\phi}p_t}{m}, \label{eq:dphi-dtau} \\
\frac{dp_t}{d\tau} &= -\frac{1}{m}\frac{\partial\mathcal{H}}{\partial t} = 0, \label{eq:dpt-dtau} \\
\frac{dp_r}{d\tau} &= -\frac{1}{m}\frac{\partial\mathcal{H}}{\partial r}, \label{eq:dpr-dtau} \\
\frac{dp_\phi}{d\tau} &= -\frac{1}{m}\frac{\partial\mathcal{H}}{\partial\phi} = 0. \label{eq:dpphi-dtau}
\end{align}
The $1/m$ factor converts covariant momentum to four-velocity in proper-time evolution;
geodesic motion remains mass-independent, consistent with the equivalence principle.
Note that $p_t = -E$ and $p_\phi = L_z$ are conserved for geodesic motion.

\section{Thrust Model and Equations of Motion}
\label{sec:thrust-model}

\paragraph{Notation reminder.}
Throughout this section: $m$ denotes the rocket's (decreasing) rest mass; $\mu$ is the cumulative exhaust rest mass ejected; $\delta m_{\rm nom}$ is the nominal burn-size parameter supplied to each run, while $\delta\mu$ and $\delta m_{\rm actual}$ are the exact exhaust rest mass and realized rocket mass loss, respectively (Sec.~\ref{sec:continuous-penrose}). The exhaust speed in the rocket frame is $v_e$ with Lorentz factor $\gamma_e = (1-v_e^2)^{-1/2}$. Killing energy is $E = -p_t$; specific energy is $E/m$; exhaust Killing energy is $E_{\rm ex} = -u_{{\rm ex},t}$ (Eq.~\ref{eq:Eex-def}).

\subsection{Variable-Mass Rocket Dynamics}

For a rocket with proper acceleration $a_{\rm prop}$ and exhaust velocity $v_e$, the mass evolution follows the rest-frame mass-flow relation:
\begin{equation}
\label{eq:dm-dtau}
\frac{dm}{d\tau} = -\frac{T}{v_e},
\end{equation}
where $T = m \cdot a_{\rm prop}$ is the thrust magnitude.

With $v_e$ denoting the exhaust speed in the rocket rest frame and Lorentz factor $\gamma_e = 1/\sqrt{1-v_e^2}$, the rocket mass loss per unit exhaust rest mass is $-dm = \gamma_e\,d\mu$ (accounting for both rest mass and kinetic energy in the rocket frame), i.e.\ $dm/d\tau = -\gamma_e\,d\mu/d\tau$. In the rocket rest frame, the exhaust carries spatial momentum $\gamma_e v_e\,d\mu$ per ejection event, so the thrust magnitude is $T = \gamma_e v_e\,d\mu/d\tau$. Substituting $d\mu/d\tau = (-dm/d\tau)/\gamma_e$ from the mass relation, the $\gamma_e$ factors cancel:
\begin{equation}
\label{eq:thrust-from-dm}
T = \gamma_e v_e \frac{d\mu}{d\tau} = \gamma_e v_e \cdot \frac{-dm/d\tau}{\gamma_e} = v_e\bigl(-\tfrac{dm}{d\tau}\bigr),
\end{equation}
consistent with Eq.~\eqref{eq:dm-dtau}.

The non-gravitational 4-force on the spacecraft is defined by
\begin{equation}
f^\mu := \frac{D p^\mu}{D\tau} - u^\mu\frac{dm}{d\tau} = m\,a^\mu,
\end{equation}
where $D/D\tau$ denotes the covariant derivative along the worldline and $a^\mu$ is the 4-acceleration. The force satisfies the orthogonality condition:
\begin{equation}
\label{eq:orthogonality}
u_\mu f^\mu = 0,
\end{equation}
which preserves the 4-velocity normalization $u_\mu u^\mu = -1$; the evolving mass shell $p_\mu p^\mu = -m^2(\tau)$ changes consistently through $dm/d\tau$.

\subsection{Thrust Direction}

To maintain physically meaningful thrust magnitude, we construct an orthonormal spatial basis in the rocket's rest frame via Gram-Schmidt orthogonalization in the $(t,r,\phi)$ subspace. Starting from the rocket's 4-velocity $u^\mu$ and coordinate basis vectors $\hat{r}^\mu = (0,1,0)$ and $\hat{\phi}^\mu = (0,0,1)$, we project each onto the subspace orthogonal to $u^\mu$ using $P^\mu_{\ \nu} = \delta^\mu_\nu + u^\mu u_\nu$ (where $u_\mu u^\mu = -1$). The \emph{radial} leg is:
\begin{align}
\tilde{e}_{(r)}^\mu &= P^\mu_{\ \nu}\,\hat{r}^\nu
  = \hat{r}^\mu + (u_\alpha \hat{r}^\alpha)\,u^\mu, \label{eq:gram-schmidt-r}\\
e_{(r)}^\mu &= \frac{\tilde{e}_{(r)}^\mu}{\sqrt{g_{\alpha\beta}\,\tilde{e}_{(r)}^\alpha \tilde{e}_{(r)}^\beta}}\,.
\end{align}
The \emph{azimuthal} leg additionally removes the component along the unnormalized radial leg:
\begin{align}
\tilde{e}_{(\phi)}^\mu &= P^\mu_{\ \nu}\,\hat{\phi}^\nu
  - \frac{g_{\alpha\beta}\tilde{e}_{(r)}^\alpha(P^\beta_{\ \nu}\hat{\phi}^\nu)}
  {g_{\alpha\beta}\tilde{e}_{(r)}^\alpha\tilde{e}_{(r)}^\beta}\,\tilde{e}_{(r)}^\mu, \label{eq:gram-schmidt-phi}\\
e_{(\phi)}^\mu &= \frac{\tilde{e}_{(\phi)}^\mu}{\sqrt{g_{\alpha\beta}\,\tilde{e}_{(\phi)}^\alpha \tilde{e}_{(\phi)}^\beta}}\,,
\end{align}
satisfying $u_\mu e_{(i)}^\mu = 0$ and $e_{(i)\mu}e_{(j)}^\mu = \delta_{ij}$. The unit thrust direction is
\begin{equation}
s^\mu = \sin\alpha\,e_{(r)}^\mu + \sigma\cos\alpha\,e_{(\phi)}^\mu, \quad \sigma\in\{+1,-1\},
\end{equation}
and $f^\mu = T\,s^\mu$, enforcing both $u_\mu f^\mu=0$ and $f_\mu f^\mu = +T^2$.

\subsection{Optimal Thrust Selection}
\label{sec:optimal-thrust-selection}

The azimuthal thrust sign ($\pm$) is chosen to maximize instantaneous energy gain:
\begin{equation}
\text{Choose sign}(\pm) \quad \text{to maximize} \quad -\frac{dp_t}{d\tau}.
\end{equation}
Since $E = -p_t$, this maximizes the rate of energy extraction.

\subsection{Exact 4-Momentum Conservation}
\label{sec:momentum-conservation}

For exact energy accounting, we use 4-momentum conservation rather than 
approximate thrust models. When the rocket ejects exhaust of rest mass $\delta\mu$ 
with 4-velocity $u^\mu_{\rm ex}$, conservation requires:
\begin{equation}
\label{eq:momentum-conservation}
p'_\mu = p_\mu - \delta\mu \cdot u_{{\rm ex},\mu},
\end{equation}
where $p'_\mu$ is the new rocket 4-momentum. The exhaust 4-velocity is:
\begin{equation}
\label{eq:exhaust-4vel}
u^\mu_{\rm ex} = \gamma_e(u^\mu - v_e s^\mu),
\end{equation}
where $s^\mu$ is the unit spacelike \emph{thrust direction} (the direction of the reaction force on the rocket) and $\gamma_e = 1/\sqrt{1-v_e^2}$. Since the exhaust is ejected opposite to the thrust, the minus sign ensures correct physics.
The covariant exhaust 4-velocity is $u_{{\rm ex},\mu} = g_{\mu\nu}u^\nu_{\rm ex}$.

The exhaust Killing energy is:
\begin{equation}
\label{eq:Eex-def}
E_{\rm ex} = -u_{{\rm ex},t} = -\gamma_e(u_t - v_e s_t).
\end{equation}
Here $E_{\rm ex}$ is the exhaust Killing energy per unit exhaust rest mass, since $u^\mu_{\rm ex}$ is a 4-velocity (unit timelike vector). \emph{Negative} $E_{\rm ex}$ is the local kinematic signature that the exhaust occupies a negative-Killing-energy state; such exhaust is confined to the ergoregion and is captured (cf.\ Sec.~\ref{sec:problem} and Ref.~\cite{grib2014negative}). Absorption of this negative Killing energy would, if backreaction were included, decrease the hole's mass parameter and rotational energy.

For continuous thrust, the rates are:
\begin{align}
\label{eq:continuous-momentum-rate}
\frac{dp_\mu}{d\tau} &= -\frac{d\mu}{d\tau}\,u_{{\rm ex},\mu},\\
\label{eq:continuous-mass-flow}
\frac{dm}{d\tau} &= -\gamma_e \frac{d\mu}{d\tau}, \qquad -dm=\gamma_e\,d\mu.
\end{align}
where $d\mu/d\tau$ is the exhaust rest-mass ejection rate. The relation between 
rocket mass loss and thrust magnitude is:
\begin{equation}
\label{eq:continuous-thrust-rate}
T = \gamma_e v_e \frac{d\mu}{d\tau} = v_e \left(-\frac{dm}{d\tau}\right).
\end{equation}
The last equality in Eq.~\eqref{eq:continuous-thrust-rate} follows directly by substituting Eq.~\eqref{eq:continuous-mass-flow} into $T=\gamma_e v_e\,d\mu/d\tau$. Similarly, substituting Eq.~\eqref{eq:exhaust-4vel} into Eq.~\eqref{eq:continuous-momentum-rate} gives
\[
\frac{dp_\mu}{d\tau}=-\gamma_e\frac{d\mu}{d\tau}(u_\mu-v_e s_\mu).
\]
Taking the $t$ component and using $E=-p_t$ and Eq.~\eqref{eq:Eex-def}, one obtains $dE/d\tau=-(d\mu/d\tau)E_{\rm ex}$. Integrating over the burn gives the continuous form of energy conservation, which in the discrete implementation is verified by:
\begin{equation}
\Delta E + \sum_i E_{{\rm ex},i} \cdot \delta\mu_i = 0,
\end{equation}
where the sum is over all thrust events. This relation is satisfied to 
residuals $< 10^{-9}$ in our simulations, as it follows from the exact 4-momentum conservation imposed at each thrust event.

\subsection{Full Equations of Motion}

Between thrust events, the spacecraft follows geodesic Hamiltonian evolution (Eqs.~\ref{eq:dr-dtau}--\ref{eq:dpphi-dtau}). In continuous-thrust mode, we add the covariant momentum source implied by exact 4-momentum conservation with the exhaust:
\begin{equation}
\left(\frac{dp_\mu}{d\tau}\right)_{\rm thrust} = -\frac{d\mu}{d\tau}\,u_{{\rm ex},\mu},
\qquad
\frac{dm}{d\tau} = -\gamma_e \frac{d\mu}{d\tau},
\end{equation}
with $u_{{\rm ex},\mu} = \gamma_e(u_\mu - v_e s_\mu)$ from Eq.~\ref{eq:exhaust-4vel}. The full forced system can be written compactly as
\begin{equation}
\frac{dp_\mu}{d\tau} = \left(\frac{dp_\mu}{d\tau}\right)_{\rm geo} + \left(\frac{dp_\mu}{d\tau}\right)_{\rm thrust},
\end{equation}
where $(dp_t/d\tau)_{\rm geo}=(dp_\phi/d\tau)_{\rm geo}=0$ and $(dp_r/d\tau)_{\rm geo}=-(1/m)\,\partial\mathcal{H}/\partial r$ (Eq.~\ref{eq:dpr-dtau}). Position updates use $p^\mu = g^{\mu\nu}p_\nu$ and $dx^\mu/d\tau = p^\mu/m$. In impulsive-thrust mode, we apply the discrete version of Eq.~\ref{eq:momentum-conservation} at the burn event and then resume geodesic evolution.

\subsection{Throttle Function}
\label{sec:throttle-function}

We activate thrust only inside the \emph{extraction zone} where Penrose extraction ($E_{\rm ex} < 0$) is achievable. We compute the extraction limit radius $R_{\rm extraction}$ dynamically as the maximum radius where $E_{\rm ex} < 0$ is geometrically possible given the current orbit parameters. We implement this using a smooth throttle function $\chi(r)$:
\begin{equation}
\chi(r) = 
\begin{cases}
0, & r \leq r_{\rm in}\;\text{or}\; r \geq r_{\rm out} \\
S\!\left(\frac{r - r_{\rm in}}{w}\right)\left[1 - S\!\left(\frac{r - r_{\rm out}}{w}\right)\right], & \text{otherwise}
\end{cases}
\end{equation}
where $r_{\rm in} := r_+ + \epsilon_{\rm in}$ and $r_{\rm out} := R_{\rm extraction} - \epsilon_{\rm out}$ enforce safety margins near the horizon and the extraction boundary; $S(x)$ is a quintic smoothstep providing $C^2$-continuous ramp transitions, and $w = 0.15\,(r_{\rm out} - r_{\rm in})$ sets the ramp width proportional to the extraction zone thickness. We use $\epsilon_{\rm in} = 5\times 10^{-3}M$ and $\epsilon_{\rm out} = 5\times 10^{-3}M$.

We determine $R_{\rm extraction}$ by scanning radii outward from $r_+$ and, at each radius, scanning over all thrust directions $(\alpha, \sigma)$ to find the minimum achievable $E_{\rm ex}$. The largest radius at which $\min_{\alpha,\sigma} E_{\rm ex} < 0$ defines $R_{\rm extraction}$. This depends on the current orbit parameters $(E, L_z)$, the current mass $m$, exhaust velocity $v_e$, and spin $a$; for single-impulse burns (applied at periapsis during geodesic infall), these coincide with the initial values $(E_0, L_{z,0})$, while for continuous thrust $R_{\rm extraction}$ is recomputed from the instantaneous state. Note that $R_{\rm extraction}$ represents the region where negative-energy exhaust is \emph{kinematically accessible for at least one thrust direction} (the one minimizing $E_{\rm ex}$); the actual controller (PID for $\alpha$, greedy sign choice for $\sigma$) does not guarantee $E_{\rm ex} < 0$ at every active step within this region.

The smoothstep function is:
\begin{equation}
S(x) = \begin{cases}
0 & x \leq 0 \\
6x^5 - 15x^4 + 10x^3 & 0 < x < 1 \\
1 & x \geq 1
\end{cases}
\end{equation}

\paragraph{Continuous-thrust amplitude.}
The proper acceleration is modulated spatially by the throttle function:
\begin{equation}
\label{eq:aprop-throttle}
a_{\rm prop}(r) = a_{\max}\,\chi(r),
\end{equation}
where $a_{\max}$ is the nominal maximum proper acceleration (a simulation parameter; we use $a_{\max} = 0.15$ in escape-optimized runs and $a_{\max} = 0.35$ in capture-mode runs, in units where $M=c=1$). Combining with the mass-flow relation (Sec.~\ref{sec:continuous-penrose}), the exhaust rest-mass ejection rate is
\begin{equation}
\label{eq:dmu-throttle}
\frac{d\mu}{d\tau} = \frac{m\,a_{\max}\,\chi(r)}{\gamma_e\,v_e},
\end{equation}
and the thrust magnitude reads
\begin{equation}
\label{eq:T-throttle}
T(r) = \gamma_e v_e \frac{d\mu}{d\tau} = m\,a_{\max}\,\chi(r),
\end{equation}
which decreases with fuel consumption ($m\to m_f$) and vanishes outside the extraction zone ($\chi=0$ for $r\notin[r_{\rm in},r_{\rm out}]$). Burning continues while $\chi(r)>0$ and $m > m_{\rm min}$ (fuel budget); we use $m_{\rm min} = 0.1\,m_0$. Equations~\eqref{eq:aprop-throttle}--\eqref{eq:T-throttle} close the continuous-thrust ODE system.

\subsection{Steering and Protocol Summary}
\label{sec:protocol-summary}

Our Monte Carlo experiments use explicit steering prescriptions rather than solving a global optimal-control problem. For reproducibility, we summarize the protocol used in both impulse and continuous-thrust simulations:

\begin{itemize}
\item \textbf{Thrust modes.} (i)~\emph{Periapsis impulse}: a single burn is executed at the first periapsis passage, expending a prescribed nominal burn-size parameter $\delta m_{\rm nom}$ (a labeling parameter; see Sec.~\ref{sec:continuous-penrose} for the label/actual distinction). (ii)~\emph{Continuous thrust}: thrust is applied continuously while the throttle $\chi(r)$ is nonzero (Sec.~\ref{sec:throttle-function}) until the fuel budget is expended or a termination event occurs (escape/capture/constraint failure).
\item \textbf{Thrust direction selection (impulse mode).} For periapsis-impulse burns, the thrust direction $(\alpha, \sigma)$ is jointly optimized by exhaustive scan over 37 uniformly spaced angles $\alpha\in[-\pi/2,\pi/2]$ with both azimuthal signs $\sigma\in\{+1,-1\}$ (74 candidate directions total), selecting the direction that minimizes the exhaust Killing energy $E_{\rm ex}$ subject to the post-impulse unbound-energy filter $E > m_{\rm after}$ (necessary but not sufficient for escape; actual escape is verified by trajectory integration, Sec.~\ref{sec:continuous-penrose}). This greedy direction selection is the ``minimize~$E_{\rm ex}$'' prescription referenced throughout.
\item \textbf{Thrust direction selection (continuous mode).} The azimuthal sign $\sigma\in\{+1,-1\}$ is chosen greedily at each step to maximize instantaneous Killing-energy gain ($-dp_t/d\tau$). The radial steering angle $\alpha$ is set by a PID controller targeting $r_{\rm set}\sim 1.5M$ to extend ergosphere dwell time while avoiding horizon approach (Sec.~\ref{sec:radial-controller}).
\item \textbf{Exact energy accounting.} Burn updates use exact 4-momentum conservation with an explicit exhaust 4-velocity (Sec.~\ref{sec:momentum-conservation}). Exhaust with $E_{\rm ex}<0$ cannot reach spatial infinity, since negative-energy worldlines cannot reach the region where $\partial_t$ is timelike.
\end{itemize}

\textbf{Why both strategies?} A simple optimal-allocation argument (Sec.~\ref{sec:optimality}) shows that, for a fixed trajectory and exhaust-direction schedule, concentrating all ejection at the point where $E_{\rm ex}$ is most negative maximizes energy extraction. Single-impulse thrust approximates this optimum. Continuous thrust is studied to (i)~provide numerical evidence for the path-averaging penalty predicted by this argument, and (ii)~connect to realistic rocket operation where truly instantaneous thrust is an idealization.

\section{Numerical Methods}
\label{sec:numerical-methods}

We work in geometric units $G=c=M=1$ throughout; exhaust velocities quoted as fractions of $c$ (e.g., $v_e = 0.95c$) are dimensionless with $c=1$ implicit. We adopt the test-particle limit ($m_0 \ll M$) so that backreaction and self-force are negligible. We normalize the spacecraft's initial rest mass to $m_0=1$; the initial specific energy $E_0 = -p_t^{(0)}/m_0$ and specific angular momentum $L_{z,0} = p_\phi^{(0)}/m_0$ are therefore dimensionless. These are independent normalizations: $G=c=M=1$ sets the spacetime scale, while $m_0=1$ is a bookkeeping convention for the spacecraft sector; the physical test-particle assumption $m_0/M \to 0$ is maintained throughout. As mass evolves, $p_t$ and $p_\phi$ are the total (not specific) covariant momenta; we track the specific energy $E/m = -u_t$ to diagnose escape ($E/m > 1$ at large radius).


\subsection{Solver Selection}

For parameter sweeps (single-impulse trajectories), we use SciPy's DOP853 integrator (8th-order Dormand-Prince) with adaptive step control, achieving near-conservation of Killing quantities during geodesic phases (mass-shell residual $<10^{-9}$; energy-balance residual $<10^{-9}$; see Appendix~\ref{app:diagnostics}). For continuous-thrust simulations, we employ classical fourth-order Runge-Kutta (RK4) with fixed timestep $\Delta\tau = 0.005M$ and mass-shell projection after each step (enabling transparent debugging of near-horizon physics and thrust events). The projection modifies $p_r$ to satisfy $g^{\mu\nu}p_\mu p_\nu + m^2 = 0$ while preserving $E = -p_t$.

\textbf{Integrator validation.} We verified that DOP853, RK45, and Radau methods yield identical trajectory outcomes (escape vs.\ capture) for all tested initial conditions. Convergence tests confirm that success rates are stable across timesteps $\Delta\tau \in [0.002, 0.02]M$ when projection is applied (see Appendix~\ref{app:diagnostics}).

\subsection{Termination Events}

We terminate integration upon: horizon approach ($r < r_+ + 0.02M$), mass depletion ($m < 0.1$), constraint violation ($|\mathcal{C}| > 10^{-1}$), or escape ($r > 50M$ with $dr/d\tau > 0$ and $E/m > 1$).

\subsection{Initial Conditions}

Initial conditions must represent a valid geodesic. Given $(r_0, E_0, L_{z,0}, m_0)$ with $r_0 = 15M$ (outside the ergosphere, allowing infall to be geodesic), we set:
\begin{align}
p_t^{(0)} &= -E_0, \\
p_\phi^{(0)} &= L_{z,0}, \\
p_r^{(0)} &= -\sqrt{\frac{-(g^{tt}p_t^{(0)2} + 2g^{t\phi}p_t^{(0)}p_\phi^{(0)} + g^{\phi\phi}p_\phi^{(0)2} + m_0^2)}{g^{rr}}},
\end{align}
where the negative sign for $p_r^{(0)}$ indicates an initially infalling trajectory (moving toward smaller $r$).

For the initial state to be physical (i.e., within the allowed region for geodesic motion), we require:
\begin{equation}
g^{tt}p_t^2 + 2g^{t\phi}p_t p_\phi + g^{\phi\phi}p_\phi^2 + m^2 < 0,
\end{equation}
ensuring $p_r^2 > 0$ (real radial momentum).

\subsection{Radial Controller}
\label{sec:radial-controller}

To extend ergosphere duration, a PID controller adjusts the radial steering angle: $\alpha = \text{clamp}[k_p(r_{\rm set} - r) - k_d\dot{r} + k_i \int (r_{\rm set} - r)\,d\tau]$, with gains $k_p = 50$, $k_d = 30$, $k_i = 5$, maximum angle $\alpha_{\rm max} = 60^\circ$, and target radius $r_{\rm set} \sim 1.5M$.

\section{Numerical Results}
\label{sec:validation}

We present systematic numerical experiments characterizing the Penrose process. The main mission family is initially unbound equatorial prograde flybys; broader scans also include nearby bound and retrograde initial conditions to map failure boundaries. Our central finding is that \emph{successful energy extraction with spacecraft escape is statistically rare in broad, untuned scans}, occurring in a narrow region of parameter space, but is numerically verified to the stated tolerances when achieved. In the focused Gaussian sweet-spot prior, success rates can reach $\sim$70\% (Sec.~\ref{sec:thrust-params}). All simulations use geometric units $G = c = M = 1$.

\textbf{Scope notes.} Most ensemble statistics in this section are for \emph{single periapsis impulses}; continuous-thrust trajectories are analyzed separately in Sec.~\ref{sec:thrust-comparison} on representative matched trajectories only. All quoted thresholds (spin, velocity onset, success rates) are specific to the equatorial prograde-flyby family, the Gaussian sweet-spot prior, and the steering prescriptions described in Sec.~\ref{sec:protocol-summary}; different orbit families, priors, or steering laws may shift these thresholds.

\begin{figure*}[t]
\centering
\includegraphics[width=\textwidth]{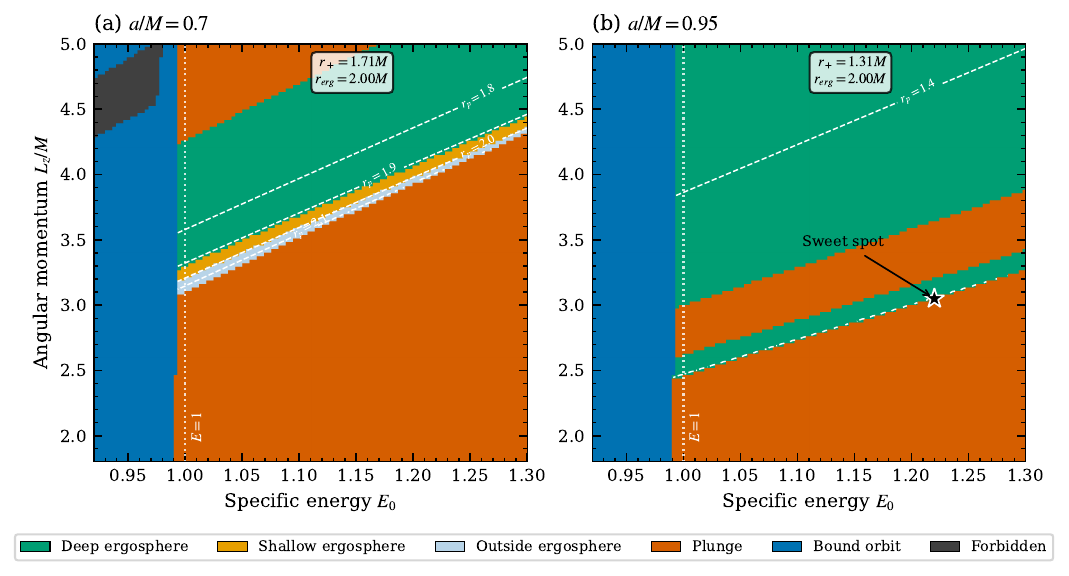}
\vspace{-0.9cm}
\caption{Orbit classification in $(E_0, L_z)$ parameter space comparing moderate and high spin. (a)~$a/M = 0.70$: Six classification regions are present---forbidden (dark gray, inaccessible configurations), bound (blue, $E_0 < 1$), plunge (red, no turning point), outside ergosphere (light blue, $r_{\rm peri} > r_{\rm erg}$), shallow ergosphere (orange, outer 30\% of ergosphere), and deep ergosphere (green, deep-ergosphere diagnostic region). The deep ergosphere region is negligible due to the narrow ergosphere width ($0.29M$). (b)~$a/M = 0.95$: Higher spin dramatically expands the deep ergosphere region (43.7\% of the plotted $(E_0, L_z)$ domain) at the expense of the outside region (now absent) and the shallow ergosphere band (greatly reduced). The wider ergosphere ($0.69M$) and stronger frame-dragging enable more trajectories to reach the deep-ergosphere region. Note: the actual $E_{\rm ex}<0$ boundary ($R_{\rm extraction}$) is state-dependent and computed dynamically (Sec.~\ref{sec:throttle-function}); this fixed-depth diagnostic cut provides a geometric proxy, not the physical extraction condition. Dashed white contours show constant periapsis $r_p$; vertical dotted line marks $E_0 = 1$ (bound/unbound threshold). Star marker in panel (b) indicates the sweet spot $(E_0, L_z) = (1.22, 3.05)$.}
\label{fig:orbit-classification}
\end{figure*}

\textbf{Organization of results.} Our analysis proceeds as follows: Sec.~\ref{sec:numerical-validation} validates numerical accuracy; Sec.~\ref{sec:extraction-window} identifies the extraction window using orbit classification (Fig.~\ref{fig:orbit-classification}); Sec.~\ref{sec:ensemble} quantifies success rates via Monte Carlo sampling (Tables~\ref{tab:broad-outcomes}--\ref{tab:focused-outcomes}, Fig.~\ref{fig:ensemble-statistics}); Sec.~\ref{sec:thrust-comparison} compares thrust strategies (Table~\ref{tab:thrust-comparison}, Fig.~\ref{fig:thrust-comparison}); Sec.~\ref{sec:spin-dependence} examines spin dependence (Fig.~\ref{fig:spin-dependence}); Sec.~\ref{sec:thrust-params} characterizes the velocity onset threshold (Tables~\ref{tab:thrust-params}--\ref{tab:thrust-params-99}, Figs.~\ref{fig:thrust-sensitivity}--\ref{fig:ultrarel-saturation}); and Sec.~\ref{sec:discussion} synthesizes these findings. The optimality argument (Sec.~\ref{sec:optimality}) motivates the thrust-strategy comparison by establishing, a priori, why single-impulse ejection should outperform continuous thrust. Broad rarity claims derive from the grid/LHS sweeps (Tables~\ref{tab:broad-outcomes}--\ref{tab:focused-outcomes}), threshold claims from the focused Gaussian scans (Tables~\ref{tab:thrust-params}--\ref{tab:thrust-params-99}), and continuous-thrust comparisons from representative matched trajectories (Table~\ref{tab:thrust-comparison}, Fig.~\ref{fig:thrust-comparison}).

\subsection{Numerical Validation}
\label{sec:numerical-validation}

Our implementation preserves the mass-shell constraint to $|\mathcal{C}|_{\rm max} < 10^{-9}$ (single-impulse) and $< 10^{-6}$ (continuous thrust with projection). We verify future-directedness ($u^t > 0$) and timelike normalization throughout. Energy conservation via exact 4-momentum conservation yields residuals $< 10^{-9}$. Detailed diagnostics are provided in Appendix~\ref{app:diagnostics}.

\subsection{Extraction Window in Parameter Space}
\label{sec:extraction-window}

We require \emph{prograde flyby orbits} satisfying: (1) $E_0 > 1$ (unbound), (2) periapsis $r_{\rm peri} < r_{\rm erg} = 2M$ (ergosphere penetration), and (3) exhaust direction chosen to minimize $E_{\rm ex}$ (in the successful sampled cases, the minimizing direction is locally retrograde and carries sufficiently negative angular momentum $L_{z,\rm ex} < 0$ to permit $E_{\rm ex} < 0$).

For diagnostics and visualization (Figs.~\ref{fig:orbit-classification} and \ref{fig:spin-dependence}), we further classify ergosphere-penetrating flybys by periapsis depth. Let $\Delta r_{\rm erg} := r_{\rm erg} - r_+$ denote the equatorial ergosphere thickness. We call a trajectory \emph{shallow} if its periapsis lies in the outer 30\% of the ergosphere, $r_{\rm peri} > r_{\rm erg} - 0.3\,\Delta r_{\rm erg}$, and \emph{deep} otherwise; deep-ergosphere flybys correspond to the diagnostic region where exhaust with sufficiently negative angular momentum ($L_{z,\rm ex} < 0$) can most robustly attain $E_{\rm ex}<0$, though the actual extraction boundary $R_{\rm extraction}$ depends on the instantaneous orbital state (Sec.~\ref{sec:throttle-function}).

For $a/M = 0.95$, we find the viable region to be remarkably narrow: specific energy $E_0 \in [1.15, 1.35]$ and specific angular momentum $L_z \in [2.9, 3.3]$, yielding periapsis at $r_{\rm peri} \approx 1.5M$. Orbits outside this window fail in different ways: initially bound conditions ($E_0 < 1$) remain bound under our baseline burns; low-$L_z$ cases are captured; high $L_z$ places periapsis outside the ergosphere; and retrograde ($L_z < 0$) cases do not yield extraction-with-escape in the sampled domain. Figure~\ref{fig:orbit-classification} shows how orbit classification depends critically on spin: at $a/M = 0.70$, the deep ergosphere region is negligible (confined to a thin band), while at $a/M = 0.95$, it expands to dominate 44\% of the plotted $(E_0, L_z)$ diagnostic domain (Fig.~\ref{fig:orbit-classification}b).

\subsection{Statistical Ensemble Analysis}
\label{sec:ensemble}

To quantify the rarity of successful extraction with escape, we performed systematic grid and Latin Hypercube sampling (LHS) studies over the parameter space $(E_0, L_z, a)$.

Unless stated otherwise, the grid/LHS sweeps in this section use a \emph{single impulsive burn applied at periapsis}. The broad scan domain ($E_0 \in [0.95, 2.0]$, $L_z \in [-3.0, 6.0]$) intentionally includes initially bound ($E_0 < 1$) and retrograde ($L_z < 0$) conditions to map the failure boundaries around the mission-relevant prograde unbound family, not because those conditions belong to the mission family itself. Continuous-thrust trajectories are analyzed separately in Sec.~\ref{sec:thrust-comparison}.

\textbf{Statistical methodology.} We use deterministic grids for systematic exploration and Latin Hypercube sampling (LHS) for Monte Carlo validation. Confidence intervals for proportions use the Clopper-Pearson exact method\footnote{C.~J. Clopper and E.~S. Pearson, Biometrika \textbf{26}, 404 (1934).}. For deterministic grid sweeps, these intervals quantify uncertainty with respect to the implied uniform measure over the chosen finite domain, not ordinary random-sampling uncertainty. Efficiency uncertainties were cross-checked via BCa bootstrap with $10^4$ resamples\footnote{B.~Efron, J.\ Am.\ Stat.\ Assoc.\ \textbf{82}, 171 (1987).}; reported efficiency values are sample means over successful escapes. Integration failures (11.2\%) are conservatively counted as captures (Table~\ref{tab:conservative-rates}).

\begin{figure}[t]
\centering
\includegraphics[width=0.9\columnwidth]{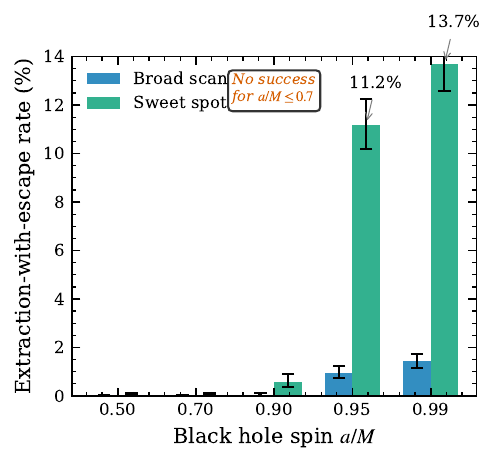}
\vspace{-0.5cm}
\caption{Extraction-with-escape success rate versus black hole spin from systematic parameter sweep: broad scan ($E_0 \in [0.95, 2.0]$, $L_z \in [-3.0, 6.0]$; blue, $N = 6{,}400$ per spin) and rectangular sweet-spot region (uniform grid, $E_0 \in [1.1, 1.4]$, $L_z \in [2.5, 3.8]$; green, $N = 3{,}600$ per spin). Error bars show 95\% Clopper-Pearson exact confidence intervals. No successful extractions were observed for $a/M \leq 0.7$ across 12{,}800 trajectories. Focusing on the sweet spot yields amplification factors of 19$\times$ at $a/M = 0.9$, 12$\times$ at $a/M = 0.95$, and 10$\times$ at $a/M = 0.99$, demonstrating the essential role of precise initial condition tuning.}
\label{fig:ensemble-statistics}
\end{figure}

\subsubsection{Broad Parameter Study}

We performed systematic parameter sweeps across a broad $(E_0, L_z)$ domain for five spin values: $E_0 \in [0.95, 2.0]$, $L_z \in [-3.0, 6.0]$, using an $80 \times 80$ grid (6{,}400 samples per spin). We choose this domain to encompass all orbits from marginally bound ($E_0 \approx 1$) to moderately relativistic ($E_0 = 2$), and from strongly retrograde to strongly prograde angular momentum, thereby encompassing the regime where ergosphere-penetrating flybys are expected for the spins in our range. We emphasize that success rates are \emph{conditional} on this finite sampling domain; since $E_0$ and $L_z$ range over non-compact intervals, the absolute probability depends on the chosen domain. Our purpose is to locate the viable extraction region and quantify its relative extent, not to assign an unconditional probability. Table~\ref{tab:broad-outcomes} summarizes the results.

\begin{table}[htpb]
\centering
\caption{Broad parameter sweep results ($N = 6{,}400$ samples per spin, $E_0 \in [0.95, 2.0]$, $L_z \in [-3.0, 6.0]$, $v_e = 0.95c$, $\delta m_{\rm nom} = 0.20$). All confidence intervals use the Clopper-Pearson exact method. The success rate quantifies trajectories achieving extraction-with-escape (Eq.~\ref{eq:penrose-success}): $\Delta E > 0$ and escape to infinity. Integration uses DOP853 (8th-order Dormand-Prince) with tolerances rtol$=10^{-9}$, atol$=10^{-11}$.}
\begin{tabular}{lcccl}
\hline
$a/M$ & $N$ & Escape & Success & 95\% CI \\
\hline
0.50 & 6400 & 8.11\% & 0.00\% & $[0.00\%, 0.06\%]$ \\
0.70 & 6400 & 8.34\% & 0.00\% & $[0.00\%, 0.06\%]$ \\
0.90 & 6400 & 13.94\% & 0.03\% & $[0.00\%, 0.11\%]$ \\
0.95 & 6400 & 19.52\% & 0.95\% & $[0.73\%, 1.22\%]$ \\
0.99 & 6400 & 27.22\% & 1.41\% & $[1.13\%, 1.73\%]$ \\
\hline
\end{tabular}

\label{tab:broad-outcomes}
\end{table}

Because integration failures (11.2\% of samples) are conservatively counted as captures (Appendix~\ref{app:diagnostics}), the reported success rates throughout this section are conservative lower estimates; exact percentages near marginal thresholds should be interpreted accordingly.

\subsubsection{Spin Threshold Analysis}

To precisely characterize the spin dependence, we performed a sweep over 14 spin values: $a/M = 0.80, 0.81, \ldots, 0.88$ plus $a/M \in \{0.89, 0.90, 0.92, 0.95, 0.99\}$, with 10,000 LHS samples per spin in the focused region. Table~\ref{tab:spin-threshold} shows the results.

\begin{table}[htpb]
\centering
\caption{Spin threshold analysis ($N = 10{,}000$ LHS samples per spin in rectangular focused region: $E_0 \in [1.1, 1.3]$, $L_z \in [2.5, 3.5]$; single periapsis-impulse burns, $v_e = 0.95c$, $\delta m_{\rm nom} = 0.3$). For compactness, the row $a/M \leq 0.88$ aggregates nine spins $a/M = 0.80, 0.81, \ldots, 0.88$ (90,000 samples total). No successful extractions for $a/M \leq 0.88$ and nonzero rates at $a/M = 0.89$ constrain the practical threshold to $0.88 < a_{\rm crit}/M \lesssim 0.89$.}
\begin{tabular}{lccl}
\hline
$a/M$ & $N$ & Success Rate & 95\% CI \\
\hline
$\leq 0.88$ & 90,000 & 0.00\% & $[0.00\%, 0.004\%]$ \\
0.89 & 10,000 & 0.09\% & $[0.04\%, 0.17\%]$ \\
0.90 & 10,000 & 0.30\% & $[0.20\%, 0.43\%]$ \\
0.92 & 10,000 & 2.47\% & $[2.18\%, 2.79\%]$ \\
0.95 & 10,000 & 8.24\% & $[7.71\%, 8.80\%]$ \\
0.99 & 10,000 & 10.92\% & $[10.32\%, 11.55\%]$ \\
\hline
\end{tabular}
\label{tab:spin-threshold}
\end{table}

Across this spin threshold sweep (140,000 additional LHS trajectories, separate from the main 104,000 in Table~\ref{tab:simulation-accounting}), \emph{no successful Penrose extractions with escape} were observed for $a/M \leq 0.88$. Combined with the onset of successful extraction at $a/M = 0.89$, this constrains the practical spin threshold for rocket-propelled Penrose extraction.

\subsubsection{Focused Sweet-Spot Study}

We define the ``sweet spot'' as the region of $(E_0, L_z)$ space corresponding to deep-ergosphere flybys with escape-capable angular momentum, identified from the orbit classification in Fig.~\ref{fig:orbit-classification}. Physically, this is the intersection of two constraints: (i) periapsis deep enough inside the ergosphere for $E_{\rm ex} < 0$ to be robustly achievable, and (ii) sufficient angular momentum for the post-burn orbit to remain unbound. For $a/M = 0.95$, this region is approximately $E_0 \in [1.1, 1.4]$, $L_z \in [2.5, 3.8]$, centered near $(E_0, L_z) \approx (1.22, 3.05)$. We apply this \emph{fixed benchmark box} (derived at $a/M = 0.95$) at all spins in Table~\ref{tab:focused-outcomes} for comparability, not because it is optimal for each spin; a spin-optimized box would likely shift the sweet-spot center and could alter cross-spin success-rate rankings. Using a $60 \times 60$ grid (3{,}600 samples per spin) restricted to this domain yields substantially higher success rates, as expected:

\begin{table}[htpb]
\centering
\caption{Focused rectangular sweet-spot sweep results (uniform grid, $N = 3{,}600$ samples per spin, $E_0 \in [1.1, 1.4]$, $L_z \in [2.5, 3.8]$, $v_e = 0.95c$, $\delta m_{\rm nom} = 0.20$; single periapsis-impulse burns). The $\sim$12$\times$ increase in extraction-with-escape success rate ($0.95\% \to 11.19\%$ for $a/M = 0.95$) and $\sim$10$\times$ increase ($1.41\% \to 13.69\%$ for $a/M = 0.99$) confirm that successful extraction-with-escape depends critically on precise tuning of initial conditions. (Note: rates differ from Table~\ref{tab:spin-threshold} due to different sampling domains and methods. The Gaussian sweet-spot prior used in Sec.~\ref{sec:thrust-params} employs a different sampling method.)}
\begin{tabular}{lcccc}
\hline
$a/M$ & $N$ & Escape & Success & 95\% CI \\
\hline
0.50 & 3600 & 0.00\% & 0.00\% & $[0.00\%, 0.10\%]$ \\
0.70 & 3600 & 2.78\% & 0.00\% & $[0.00\%, 0.10\%]$ \\
0.90 & 3600 & 18.64\% & 0.58\% & $[0.36\%, 0.89\%]$ \\
0.95 & 3600 & 26.33\% & \textbf{11.19\%} & $[10.18\%, 12.27\%]$ \\
0.99 & 3600 & 65.14\% & \textbf{13.69\%} & $[12.59\%, 14.86\%]$ \\
\hline
\end{tabular}

\label{tab:focused-outcomes}
\end{table}
The $\sim$12$\times$ amplification in success rate (see Figure~\ref{fig:ensemble-statistics}) confirms that precise initial condition tuning is essential.

\subsection{Optimality of Single-Impulse Extraction}
\label{sec:optimality}

Before comparing thrust strategies numerically, we establish a theoretical prediction: for a \emph{fixed realized trajectory and exhaust-direction schedule}, hence a fixed profile $E_{\rm ex}(\tau)$ along the worldline, the total energy change is:
\begin{equation}
\Delta E = -\int_0^{\tau_f} E_{\rm ex}(\tau) \cdot \frac{d\mu}{d\tau} \, d\tau,
\label{eq:energy-integral}
\end{equation}
where $E_{\rm ex}(\tau)$ is the exhaust Killing energy and $d\mu/d\tau$ is the exhaust rest-mass ejection rate (here $\mu$ denotes expelled rest mass, not a spacetime index). For fixed total exhaust mass and fixed $E_{\rm ex}(\tau)$ profile, maximizing $\Delta E$ requires concentrating ejection where $E_{\rm ex}$ is most negative.

Define normalized mass flow $\rho(\tau) = (1/\delta\mu_{\rm total})(d\mu/d\tau)$ with $\int \rho\,d\tau = 1$. Then $\Delta E = -\delta\mu_{\rm total} \langle E_{\rm ex} \rangle_\rho$, where the path-average satisfies $\langle E_{\rm ex} \rangle_\rho \geq \min_\tau E_{\rm ex}(\tau)$. Maximum extraction occurs when $\rho(\tau) = \delta(\tau - \tau^*)$, concentrated at the minimum, precisely a \emph{single impulse} at optimal location.

\textbf{Caveats.} (i)~The delta-function optimality assumes no upper bound on instantaneous mass flow; if a constraint $(d\mu/d\tau) \leq \dot\mu_{\max}$ is imposed, the optimizer becomes ``burn at maximum rate in an interval around $\tau^*$'' rather than a true impulse. (ii)~This argument proves optimal \emph{allocation} of a fixed fuel budget along a given worldline; global optimality over all steering laws is an optimal-control problem that we do not address here.

The exhaust Killing energy varies with radius: $E_{\rm ex} \sim 0$ at the ergosphere boundary, $\sim -0.1$ to $-0.2$ at $r \sim 1.5M$ (most favorable for escape in the sampled set), and more negative at smaller radii (but with increasing capture risk). Continuous thrust averages over this profile, suffering a path-averaging penalty. Additionally, continuous thrust modifies $(E,L_z)$ and the periapsis geometry throughout the ergosphere transit; extended burning can move the spacecraft out of the narrow escape corridor before the maneuver completes, increasing capture probability. We now illustrate this comparison with representative matched-initial-state trajectories (noting that actual fuel consumption differs between strategies; see Table~\ref{tab:thrust-comparison}). Because the impulse and continuous controllers follow different realized trajectories and consume different actual mass fractions, the numerical comparison below \emph{illustrates} the path-averaging penalty rather than constituting a controlled test of the fixed-worldline optimality result.

\subsection{Thrust Strategy Comparison}
\label{sec:thrust-comparison}

\begin{figure*}[t]
\centering
\includegraphics[width=0.9\textwidth]{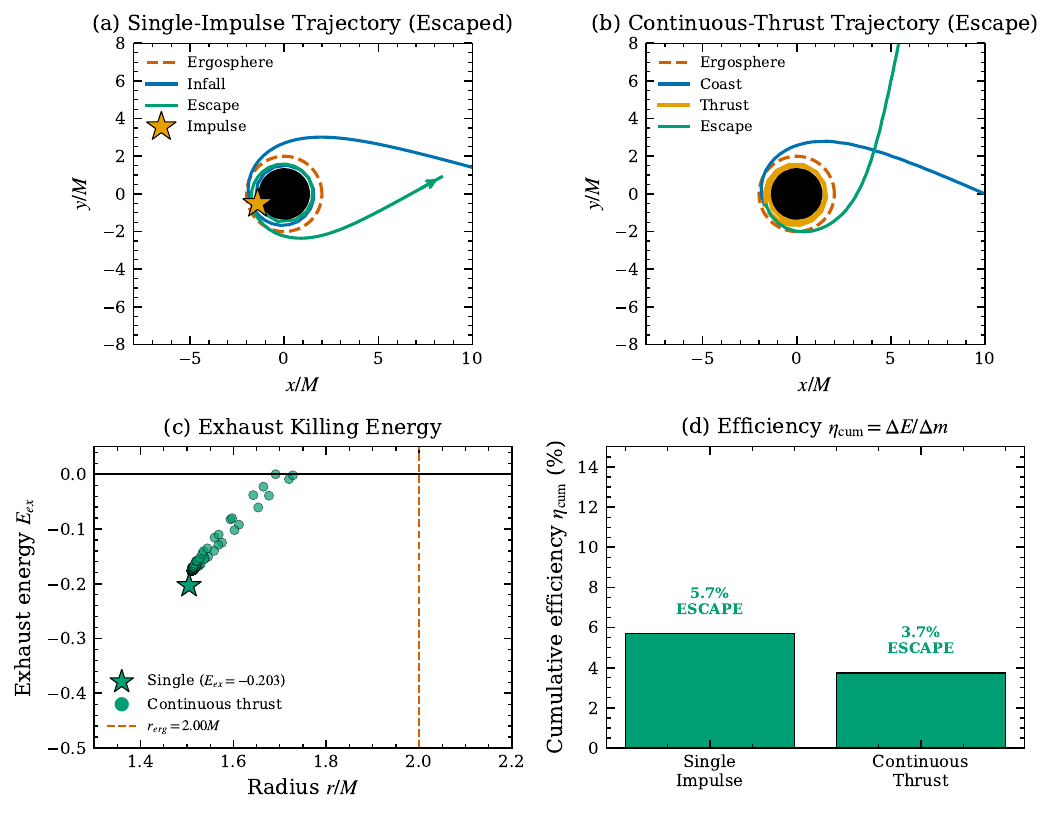}
\vspace{-0.5cm}
\caption{Thrust strategy comparison at $a/M = 0.95$, $v_e = 0.95c$. (a)~Single-impulse trajectory ($E_0 = 1.20$, $L_z = 3.0$): geodesic infall (blue), impulse at periapsis (orange star), geodesic escape (green). (b)~Continuous thrust trajectory ($E_0 = 1.20$, $L_z = 3.0$): coasting infall (blue), sustained thrust with $\sim$33{,}000 discrete events inside ergosphere (orange), escape (green). (c)~Exhaust Killing energy $E_{\rm ex}$ versus radius: large star marks the representative single-impulse event ($E_{\rm ex} \approx -0.203$); small circles show continuous thrust events. All events satisfy $E_{\rm ex} < 0$. Dashed line marks ergosphere boundary ($r_{\rm erg} = 2M$). (d)~Cumulative efficiency $\eta_{\rm cum} = \Delta E/\Delta m$ (Eq.~\ref{eq:eta-cum}) for these representative trajectories: single impulse achieves $\eta_{\rm cum} = 5.7\%$, continuous thrust $3.7\%$ (exact values in Table~\ref{tab:thrust-comparison}). Note that the Wald single-decay benchmark (20.7\%) uses energy normalization $\eta_{\rm trad} = \Delta E/E_0$ rather than propellant normalization $\eta_{\rm cum} = \Delta E/\Delta m$; the two metrics are not directly comparable.}
\label{fig:thrust-comparison}
\end{figure*}

We compare single-impulse and continuous thrust strategies to test whether concentrating ejection near periapsis---where $|E_{\rm ex}|$ is largest---outperforms distributed thrust, as predicted by the optimality argument above (Sec.~\ref{sec:optimality}). The representative escape trajectories use $(E_0, L_z) = (1.20, 3.0)$ for both the single-impulse and continuous-thrust cases shown in Fig.~\ref{fig:thrust-comparison}; Table~\ref{tab:thrust-comparison} additionally includes a continuous-capture example at $(1.20, 2.8)$ to illustrate the efficiency-capture trade-off. The actual mass fraction consumed $\delta m := (m_0 - m_f)/m_0$ differs between strategies because continuous thrust distributes ejection over the ergosphere transit while the impulse concentrates it at periapsis. Table~\ref{tab:thrust-comparison} summarizes the results.

\begin{table}[htpb]
\footnotesize
\centering
\caption{Thrust strategy comparison for $a/M = 0.95$, $v_e = 0.95c$. Each column reports a representative single trajectory at the listed initial conditions. ``Cont.\ (Esc.)'' and ``Cont.\ (Capture)'' refer to the \emph{spacecraft} outcome, not the exhaust fate. We report $\eta_{\rm cum} = \Delta E/\Delta m$ (propellant-normalized; Eq.~\ref{eq:eta-cum}), the traditional Penrose efficiency $\eta_{\rm trad} = \Delta E/E_0$ (Eq.~\ref{eq:eta-trad}), and the per-exhaust-rest-mass ratio $\Delta E/\delta\mu$ where $\delta\mu$ is computed via Eq.~\eqref{eq:dmu-exact} for the single-impulse case and via the exact differential relation $-dm = \gamma_e\,d\mu$ for continuous thrust. The single-impulse $\delta m = 0.22$ is the \emph{actual} consumed mass fraction (see Sec.~\ref{sec:continuous-penrose} for the finite-impulse discussion). These are illustrative matched trajectories; broader \emph{single-impulse} ensemble statistics appear in Tables~\ref{tab:thrust-params}--\ref{tab:thrust-params-99}. Efficiency metrics ($\eta_{\rm cum}$, $\Delta E/\delta\mu$) are computed from full-precision simulation values; displayed $\Delta E$ and $\delta m_{\rm actual}$ are independently rounded.}
\resizebox{\columnwidth}{!}{%
\begin{tabular}{lccc}
\hline
\textbf{Metric} & \textbf{Single Impulse} & \textbf{Cont.\ (Esc.)} & \textbf{Cont.\ (Capture)} \\
\hline
$(E_0, L_z)$ & $(1.20, +3.0)$ & $(1.20, +3.0)$ & $(1.20, +2.8)$ \\
$\delta m_{\rm actual}$ & 0.22 & 0.07 & 0.06 \\
$\Delta E$ & $+0.013$ & $+0.002$ & $+0.003$ \\
$\eta_{\rm cum}$ & $5.7\%$ & $3.7\%$ & $5.0\%$ \\
$\eta_{\rm trad}$ & $1.1\%$ & $0.2\%$ & $\sim$0.3\% \\
$\Delta E/\delta\mu$ & $\sim$20\% & $\sim$12\% & $\sim$16\% \\
Frac.\ events $E_{\rm ex}<0$ & 100\% & 100\% & $\sim$100\% \\
Outcome & \textbf{ESCAPE} & \textbf{ESCAPE} & CAPTURE \\
\hline
\end{tabular}}

\label{tab:thrust-comparison}
\end{table}

\textbf{Key findings:} (1) Single impulse is most efficient among the strategies studied, achieving $\eta_{\rm cum}\approx 5.7\%$ (here $\delta m_{\rm actual} = 0.22$ for $\delta m_{\rm nom} = 0.20$; see Sec.~\ref{sec:continuous-penrose} for the label/actual distinction; $\Delta E/\delta\mu \approx 20\%$ per exhaust rest mass)\footnote{The Wald single-decay bound $\eta_{\rm Wald}\approx20.7\%$ normalizes by incident particle energy ($\Delta E/E_0$), not exhaust rest mass, so the two figures are not directly comparable despite their numerical proximity. We cite $\eta_{\rm Wald}$ as a benchmark for single-decay scenarios; it is not a strict upper bound on the controlled rocket setup studied here.} by concentrating fuel at periapsis, consistent with the optimal-allocation argument of Sec.~\ref{sec:optimality}. The traditional Penrose efficiency is $\eta_{\rm trad}\approx 1.1\%$ for the single impulse ($0.2\%$ for continuous escape), well below Wald's single-decay benchmark $\eta_{\rm Wald}\approx 20.7\%$, consistent with sub-extremal spin ($a/M=0.95$) and a non-optimal orbit. (2) Continuous thrust enables escape with reduced efficiency ($\sim$60\% of single-impulse $\eta_{\rm cum}$), consistent with the path-averaging penalty. (3) Deeper periapsis increases $|E_{\rm ex}|$ but also capture probability; the continuous capture trajectory achieves higher \emph{local pre-capture} per-mass efficiency than continuous escape because all burning occurs deep in the ergosphere where $|E_{\rm ex}|$ is largest; however, this is not an extraction-with-escape success under our mission criterion (Eq.~\ref{eq:penrose-success}), since the spacecraft does not escape to infinity (cf.\ the discussion of $\mathcal{S}$ in Sec.~\ref{sec:continuous-penrose}). (4) All illustrated escape trajectories achieve $E_{\rm ex} < 0$ for all thrust events inside the ergosphere; the continuous-capture case achieves $E_{\rm ex} < 0$ for nearly all thrust events ($P(E_{\rm ex}<0)\sim 100\%$). Figure~\ref{fig:thrust-comparison} shows the trajectories reported in Table~\ref{tab:thrust-comparison}.

\subsection{Spin Dependence}
\label{sec:spin-dependence}

This section complements the spin threshold analysis of Sec.~\ref{sec:ensemble} by examining the geometric origin of the spin dependence through orbit classification maps, rather than success-rate statistics.

\begin{figure*}[t]
\centering
\includegraphics[width=\textwidth]{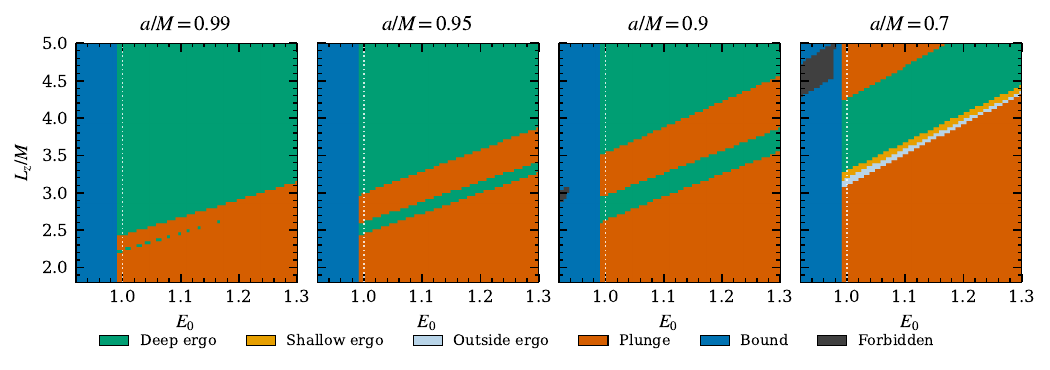}
\vspace{-0.9cm}
\caption{Spin dependence of the extraction window. Panels show orbit classification in $(E_0, L_z)$ space for $a/M = 0.99$, $0.95$, $0.9$, and $0.7$ (left to right). Color coding follows Fig.~\ref{fig:orbit-classification}: green indicates deep ergosphere flyby, orange shallow ergosphere, light blue outside ergosphere, red plunge, blue bound ($E_0 < 1$), and gray forbidden. The deep ergosphere region (deep-ergosphere diagnostic region; see Fig.~\ref{fig:orbit-classification} caption for the proxy/actual distinction) expands dramatically with increasing spin: from negligible at $a/M = 0.7$ to dominating the plotted $(E_0, L_z)$ domain at $a/M = 0.99$. The bound region (left of vertical dotted line at $E_0 = 1$) is present at all spins but not relevant for extraction in our baseline impulsive scans, since these initial conditions are bound ($E_0<1$) and do not satisfy the escape criterion. Within this prograde-flyby family, high spin ($a/M \gtrsim 0.89$) is observed to be necessary for practical extraction-with-escape. At high spin (e.g., $a/M = 0.99$), the deep ergosphere region can appear as two disjoint zones: one at moderate $L_z$ where periapsis reaches moderate ergosphere depth, and another at lower $L_z$ where orbits penetrate very close to the horizon. A thin plunge band separates them; some isolated points in the high-spin panel reflect finite grid resolution rather than truly disconnected regions. The $a/M = 0.95$ and $0.7$ panels reproduce Fig.~\ref{fig:orbit-classification} at reduced scale for side-by-side spin comparison; see that figure for detailed zone annotations.)}
\label{fig:spin-dependence}
\end{figure*}

The extraction window depends sensitively on black hole spin. In the broad parameter study (Table~\ref{tab:broad-outcomes}), no successful extractions were observed for $a/M \leq 0.7$ across 12{,}800 trajectories, while in the dedicated spin threshold analysis (Table~\ref{tab:spin-threshold}), the finer-grained result is $a_{\rm crit}/M \in (0.88, 0.89]$ using the narrower rectangular sweet-spot domain. These two findings are consistent: the broad study's coarser domain and lower spin values produce zero successes below $a/M = 0.7$, while the focused study, sampling more densely within the sweet spot, resolves the transition at higher spin. We stress that negative-energy timelike states inside the ergosphere exist for any nonzero Kerr spin; what vanishes at low spin in our analysis is the \emph{overlap} between those states and the escape-capable prograde flybys in our sampled domain, compounded by finite Monte Carlo sampling resolution. Figure~\ref{fig:spin-dependence} illustrates the contraction of the extraction window with decreasing spin.

\subsection{Thrust Parameter Sensitivity}
\label{sec:thrust-params}

Having established global conditional rarity and the sweet-spot amplification effect, we now turn to local sensitivity around the benchmark injection point. A critical finding from our systematic sweep is the sharp dependence on exhaust velocity $v_e$. To characterize this, we perform a \emph{local sensitivity study} around a fixed benchmark prior at the $a/M = 0.95$ sweet spot, not a generic survey of all orbit families or steering laws: initial conditions are sampled from a Gaussian distribution centered on $(E_0, L_z) = (1.22, 3.05)$ with standard deviations $\sigma_E = 0.03$ and $\sigma_{L_z} = 0.08$. This Gaussian represents finite-precision orbit injection: a spacecraft targeting the optimal orbit will have small uncertainties in its orbital elements, and the Gaussian widths probe sensitivity to these uncertainties. We emphasize that this distribution is anchored to the $a/M = 0.95$ sweet spot and is not a spin-optimized prior; different spins may have different optimal centers. Our velocity analysis reveals a sharp onset around $v_e \approx 0.91$--$0.92c$: below this range, success is of order 1\% or below in this Gaussian scan (Table~\ref{tab:thrust-params} shows 0.4--1\% at $v_e = 0.90c$), while above it success increases rapidly with both $v_e$ and $\delta m_{\rm nom}$. Table~\ref{tab:thrust-params} shows representative results for $a/M = 0.95$:

\begin{table}[htpb]
\centering
\caption{Thrust parameter sensitivity for $a/M = 0.95$ in the sweet-spot region, with $N = 500$ initial conditions per configuration sampled from a Gaussian distribution centered at $(E_0, L_z) = (1.22, 3.05)$ with $\sigma_E = 0.03$, $\sigma_{L_z} = 0.08$. Extraction-with-escape success rate as a function of exhaust velocity $v_e$ and nominal burn-size parameter $\delta m_{\rm nom}$ (\textbf{labeling parameter only}; the actual consumed fraction $\delta m_{\rm actual}$ exceeds this value due to the finite-impulse correction discussed in Sec.~\ref{sec:continuous-penrose}; at $v_e = 0.95c$ the discrepancy ranges from $\sim$5\% at $\delta m_{\rm nom}=0.1$ to $\sim$34\% at $\delta m_{\rm nom}=0.4$). Representative actual consumed fractions: at $v_e = 0.95c$, $\delta m_{\rm actual} \approx 0.105,\, 0.223,\, 0.361,\, 0.536$ for $\delta m_{\rm nom} = 0.1,\, 0.2,\, 0.3,\, 0.4$ respectively; at $v_e = 0.98c$, $\delta m_{\rm actual} \approx 0.105,\, 0.224,\, 0.365,\, 0.546$. A sharp onset occurs around $v_e \approx 0.91$--$0.92c$; below this threshold, success is of order 1\% or below. Peak success of \textbf{69.6\%} achieved at $v_e = 0.98c$, $\delta m_{\rm nom} = 0.4$. Corresponding cumulative efficiency trends are summarized in Fig.~\ref{fig:thrust-sensitivity}b.}
\begin{tabular}{lcccc}
\hline
$v_e/c$ & $\delta m_{\rm nom} = 0.1$ & $\delta m_{\rm nom} = 0.2$ & $\delta m_{\rm nom} = 0.3$ & $\delta m_{\rm nom} = 0.4$ \\
\hline
0.80 & 0.0\% & 0.0\% & 0.0\% & 0.0\% \\
0.90 & 0.4\% & 1.0\% & 0.4\% & 0.4\% \\
0.95 & 39.4\% & 47.8\% & 49.0\% & 63.6\% \\
0.98 & 48.0\% & 56.4\% & 63.2\% & \textbf{69.6\%} \\
\hline
\end{tabular}

\label{tab:thrust-params}
\end{table}

Exploratory results at $a/M=0.99$, now verified with independent random seeds, appear in Appendix~\ref{app:exploratory-99}; they are not part of our headline findings.

The cumulative efficiency $\eta_{\rm cum}$ (Eq.~\ref{eq:eta-cum}), computed as the sample mean over successful escapes (conditioning on $\mathcal{S}$), increases with exhaust velocity: for $a/M = 0.95$ it rises from $\eta_{\rm cum} \approx 0.1$--$0.2\%$ just above threshold ($v_e \approx 0.91$--$0.92c$) to $\eta_{\rm cum} \approx 5$--$7\%$ at $v_e = 0.98c$ (Fig.~\ref{fig:thrust-sensitivity}b). At this spin, higher $\delta m_{\rm nom}$ values increase \emph{success rates} but \emph{decrease} per-mass efficiency, presenting a design trade-off.

Within this Gaussian sweet-spot prior and our baseline family and protocol, the observed velocity threshold acts as a practical constraint on material Penrose propulsion in our simulations. Combining higher exhaust velocity ($v_e = 0.98c$) with larger nominal burn-size parameter ($\delta m_{\rm nom} = 0.4$) yields success rates of 69.6\% for $a/M = 0.95$, compared to ${\sim}39\%$ at $v_e = 0.95c$, $\delta m_{\rm nom} = 0.1$ in the same domain. Figure~\ref{fig:thrust-sensitivity} visualizes this sharp onset and the associated trade-off between success probability and per-mass efficiency.

To probe the asymptotic limit $v_e\to c$---and test whether $\eta_{\rm cum}$ continues to grow once the sharp onset is passed---we extend the same single-impulse sweet-spot study by pushing the exhaust velocity into the ultra-relativistic regime, up to $v_e = 0.99999c$ ($\gamma \approx 224$). Figure~\ref{fig:ultrarel-saturation} continues Fig.~\ref{fig:thrust-sensitivity}b and shows that the efficiency gain beyond $v_e \sim 0.99c$ is modest, approaching a plateau for this spin parameter.

\begin{figure*}[t]
\centering
\includegraphics[width=\textwidth]{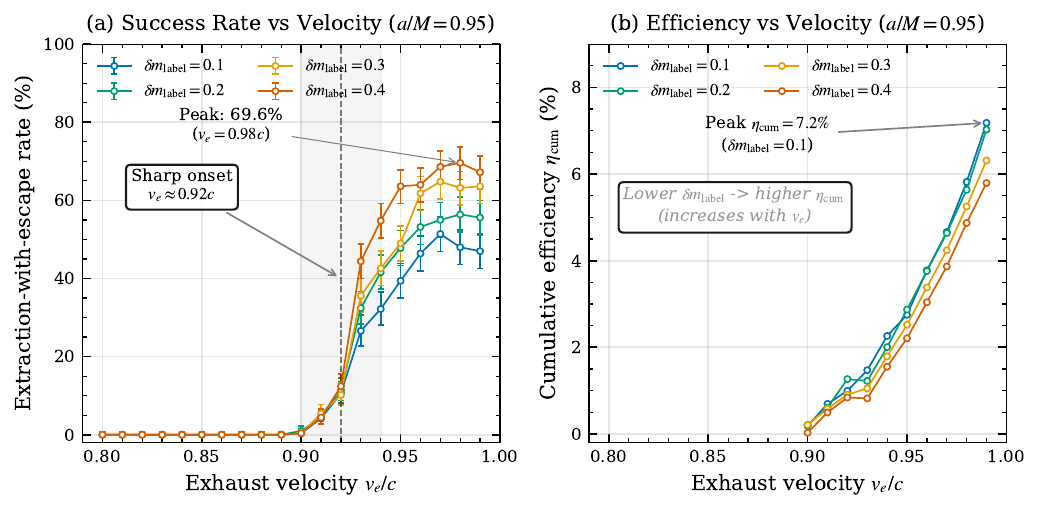}
\vspace{-0.9cm}
\caption{Thrust parameter sensitivity for $a/M = 0.95$ at the sweet spot ($E_0 = 1.22$, $L_z = 3.05$), sampled at 0.01$c$ velocity increments ($N = 500$ initial conditions per configuration from a Gaussian at $(1.22, 3.05)$ with $\sigma_E = 0.03$, $\sigma_L = 0.08$). (a)~Extraction-with-escape success rate versus $v_e$ for four $\delta m_{\rm nom} \in \{0.1, 0.2, 0.3, 0.4\}$ (labeling parameters only; $\delta m_{\rm actual}$ exceeds these values---see Table~\ref{tab:thrust-params} caption for actual consumed fractions) with 95\% Clopper-Pearson confidence intervals. Sharp onset near $v_e \approx 0.91$--$0.92c$; peak success 69.6\% at $v_e = 0.98c$, $\delta m_{\rm nom} = 0.4$. (b)~Cumulative efficiency $\eta_{\rm cum}$ (Eq.~\ref{eq:eta-cum}), sample mean over successful escapes only (bins with zero successes are omitted, as the conditional mean is undefined there). Efficiency is \emph{inversely} related to $\delta m_{\rm nom}$: smaller nominal burn sizes yield higher per-mass energy gain, reaching $\eta_{\rm cum} \approx 7.2\%$ at $v_e = 0.99c$, $\delta m_{\rm nom} = 0.1$. Success rate and efficiency trends oppose each other, presenting a design trade-off. Figure~\ref{fig:ultrarel-saturation} extends panel~(b) to the ultra-relativistic limit.}
\label{fig:thrust-sensitivity}
\end{figure*}

\begin{figure}[t]
\centering
\includegraphics[width=\columnwidth]{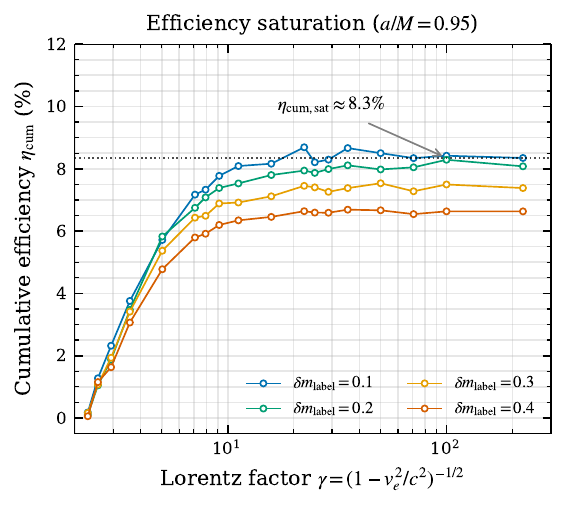}
\vspace{-0.5cm}
\caption{Continuation of Fig.~\ref{fig:thrust-sensitivity}b into the extreme-$\gamma$ regime for $a/M = 0.95$ in the same sweet-spot region (Eq.~\ref{eq:eta-cum}). All $\eta_{\rm cum}$ values are sample means over successful escapes (conditioning on $\mathcal{S}$). Exhaust velocities span $v_e = 0.90c$ ($\gamma = 2.3$) to $v_e = 0.99999c$ ($\gamma = 224$). Below $\gamma \approx 10$, $\eta_{\rm cum}$ grows approximately linearly with $\gamma$; above $\gamma \approx 20$--30, it saturates at $\eta_{\rm sat} \approx 8.3\%$, indicating diminishing returns at this spin. Lower $\delta m_{\rm nom}$ consistently yields higher $\eta_{\rm cum}$ ($\delta m_{\rm nom}$ is a nominal labeling parameter; see Table~\ref{tab:thrust-params} caption for actual consumed fractions). This saturation plateau is an empirical ceiling for our Gaussian sweet-spot prior and thrust policy. We plot against Lorentz factor $\gamma$ rather than $v_e$ to resolve the saturation regime, where fractional changes in $v_e$ become negligible (e.g., $v_e = 0.99c$ and $v_e = 0.999c$ differ by only $0.009$).}
\label{fig:ultrarel-saturation}
\end{figure}


\subsection{Discussion}
\label{sec:discussion}

Figures~\ref{fig:thrust-sensitivity} and \ref{fig:ultrarel-saturation} show two stages: a sharp onset of viable extraction near $v_e \approx 0.91$--$0.92c$, then diminishing efficiency gains as $v_e\to c$. This is consistent with the local kinematic estimate of Sec.~\ref{sec:velocity-threshold}: the Killing-energy decomposition $E_{\rm ex} = -\gamma_e(u_t - v_e s_t)$ requires $v_e > v_e^{\rm crit} := |u_t|/|s_t| \approx 0.85$--$0.92c$ (depending on periapsis depth) for negative-energy exhaust. The Monte Carlo results agree: success rates remain below ${\sim}1\%$ for $v_e \leq 0.90c$ but rise steeply for $v_e \geq 0.93c$. Beyond this onset, $\eta_{\rm cum}$ approaches a plateau for $a/M=0.95$ (Fig.~\ref{fig:ultrarel-saturation}). The behavior at $a/M = 0.99$ is discussed in Appendix~\ref{app:exploratory-99}.

In our sampled prograde-flyby family, the observed spin threshold is consistent with ergosphere geometry: at $a/M = 0.7$, the ergosphere is shallow and the parameter space allowing deep penetration with escape is negligible in our sampled domain. The 10$\times$--19$\times$ amplification from sweet-spot focusing confirms that initial condition precision dominates over other factors. We note that the velocity-onset location ($v_e \approx 0.91$--$0.92c$) has been characterized only for a single Gaussian prior width and center; varying these or using a different prior family may shift the threshold. A systematic prior-sensitivity study is left for future work.

Comparing our results with electromagnetic extraction mechanisms is instructive. The Blandford--Znajek process~\cite{blandford1977electromagnetic} achieves sustained extraction through magnetic field lines without requiring ultra-relativistic \emph{material} exhaust, helping contextualize its astrophysical prevalence. Magnetic reconnection in the ergosphere~\cite{comisso2021magnetic} similarly bypasses a material exhaust-velocity threshold by accelerating charged particles electromagnetically. Our results are consistent with the view that material-based Penrose extraction, while theoretically elegant, faces more stringent parameter requirements than electromagnetic channels in realistic astrophysical settings.

Related relativistic rocket work has examined the Oberth effect and continuous fuel ejection in Schwarzschild spacetime~\cite{pavlov2023oberth}. That problem shares the variable-mass rocket accounting used here, but it is not a direct analogue of the present Kerr calculation: Schwarzschild spacetime has no ergosphere and no exterior negative-Killing-energy states, so its energy gain is an Oberth/redshift effect rather than extraction of rotational energy. The comparison is nevertheless useful because it separates generic relativistic-rocket bookkeeping from the specifically Kerr ingredient in this paper, namely controlled exhaust entering negative-energy states in the ergoregion.

\textbf{Summary of experimental evidence.} The principal quantitative results, synthesized in Sec.~\ref{sec:conclusions}, derive from the following analyses: parameter space structure (Fig.~\ref{fig:orbit-classification}); baseline success rates (Table~\ref{tab:broad-outcomes}); spin threshold (Table~\ref{tab:spin-threshold}); sweet-spot amplification (Table~\ref{tab:focused-outcomes}, Fig.~\ref{fig:ensemble-statistics}); thrust strategy efficiency (Table~\ref{tab:thrust-comparison}, Fig.~\ref{fig:thrust-comparison}); spin dependence (Fig.~\ref{fig:spin-dependence}); velocity onset (Table~\ref{tab:thrust-params}, Fig.~\ref{fig:thrust-sensitivity}); ultra-relativistic efficiency saturation (Fig.~\ref{fig:ultrarel-saturation}); and exploratory $a/M=0.99$ results (Appendix~\ref{app:exploratory-99}).

\section{Conclusions}
\label{sec:conclusions}

We have presented a Monte Carlo characterization of the Penrose process using rocket propulsion in Kerr spacetime, framing controlled exhaust ejection as a controlled-ejection variant of the classical Penrose decay. The statistical core of this study is the periapsis-impulse ensemble (Tables~\ref{tab:broad-outcomes}--\ref{tab:thrust-params}); the continuous-thrust comparison is a secondary study of representative matched trajectories under specific implemented controllers (Table~\ref{tab:thrust-comparison}, Fig.~\ref{fig:thrust-comparison}). All constraints below apply to our equatorial prograde-flyby family under the steering prescriptions of Sec.~\ref{sec:protocol-summary}; different orbit families or steering laws may shift thresholds.

\begin{enumerate}
\item \textbf{Extraction requires fine-tuning.} At most 1.4\% of trajectories in a deliberately broad diagnostic domain ($E_0\in[0.95,2.0]$, $L_z\in[-3,6]$, which includes bound and retrograde states outside the mission family; Table~\ref{tab:broad-outcomes}) achieve extraction-with-escape success ($\Delta E > 0$ with escape to infinity, Eq.~\ref{eq:penrose-success}), with success of order 1\% only at the highest sampled spins. Under maximal tuning ($v_e = 0.98c$, $\delta m_{\rm nom} = 0.4$, corresponding to $\delta m_{\rm actual} \approx 0.55$), success rates reach 70\% for $a/M = 0.95$ (Gaussian sweet-spot prior). Exploratory results at $a/M = 0.99$ yield comparable rates (${\sim}65$--$70\%$) under independent-seed verification (Appendix~\ref{app:exploratory-99}). These represent the fine-tuned limit, not generic behavior. All reported success rates are conservative lower bounds, as integration failures (11.2\% of samples) are counted as captures; because these failures cluster near the horizon, this conservative counting may also slightly bias estimated threshold locations (Appendix~\ref{app:diagnostics}).

\item \textbf{High spin is observed to be necessary within the explored family.} Within our baseline family and protocol and sampled $(E_0,L_z)$ domain, successful extraction-with-escape is observed only at high spin: no successes occur for $a/M \leq 0.88$ (0/90{,}000 trajectories at $v_e=0.95c$, $\delta m_{\rm nom}=0.3$), constraining the practical threshold to $0.88 < a_{\rm crit}/M \lesssim 0.89$ within the present Boyer--Lindquist numerics. This is an empirical boundary of the sampled region, not a coordinate-independent or universal lower bound on Penrose extraction.

\item \textbf{Highly relativistic exhaust is empirically required within the Gaussian sweet-spot prior.} Under the Gaussian sweet-spot prior (centered at $E_0=1.22$, $L_z=3.05$) and implemented steering prescriptions, we observe a sharp empirical onset near $v_e \approx 0.91$--$0.92c$ within the present Boyer--Lindquist numerics, below which success is of order 1\% or below in the sampled prior and above which it increases rapidly. Negative exhaust Killing energy requires $v_e > v_e^{\rm crit} = |u_t|/|s_t|$ (Eq.~\ref{eq:ve-crit}), a condition that becomes satisfiable over a wide range of periapsis configurations only at sufficiently relativistic speeds. Different orbit families or steering laws may yield different critical velocities.

\item \textbf{Single impulse at periapsis is most efficient among the strategies studied.} Concentrating fuel at the point where $E_{\rm ex}$ is most negative maximizes extraction (an optimal-allocation result for a fixed worldline, exhaust-direction schedule, and fuel budget). For the specific implemented controllers and representative escape trajectories started from the same initial condition (Table~\ref{tab:thrust-comparison}), impulsive thrust achieves $\eta_{\rm cum} \approx 5.7\%$ ($\sim$20\% per exhaust rest mass; $\eta_{\rm trad}\approx 1.1\%$). Continuous thrust yields $\eta_{\rm cum} \approx 3.7\%$ for the representative matched escape trajectory and implemented controllers ($\sim$65\% of the impulse value), consistent with path-averaging and trajectory deformation under extended burning; this comparison is between specific controllers, not a general theorem about continuous thrust. Global optimality over all possible steering laws remains an open optimal-control problem (Sec.~\ref{sec:optimality}).

\item \textbf{Efficiency saturates empirically.} Over the explored range up to $v_e = 0.99999c$ ($\gamma \approx 224$), $\eta_{\rm cum}$ levels off near $8.3\%$ for $a/M = 0.95$ (Fig.~\ref{fig:ultrarel-saturation}), with diminishing returns above $v_e \approx 0.999c$. This is an empirical ceiling for the Gaussian sweet-spot prior and implemented steering prescriptions, not a theoretical upper bound on Penrose extraction efficiency.
\end{enumerate}

Our results characterize the controlled-exhaust regime of the Penrose process, a controlled-ejection variant distinct from the collisional mechanisms of Piran \textit{et al.}~\cite{piran1975high} and Leiderschneider and Piran~\cite{leiderschneider2016maximal}. The stringent requirements on spin, velocity, and orbital geometry are consistent with the comparative difficulty of material-exhaust Penrose channels relative to electromagnetic ones (Blandford--Znajek, magnetic reconnection), which are generally more astrophysically relevant for energy extraction from rotating black holes.

\textbf{Engineering summary.} Within our equatorial prograde-flyby family, Gaussian sweet-spot prior, and steering prescriptions, non-negligible success is observed only under: (i)~$a/M \gtrsim 0.89$, (ii)~$v_e \gtrsim 0.91c$, and (iii)~initial conditions tuned to $(E_0, L_z) \approx (1.2, 3.0)$. Single-impulse efficiency is $\eta_{\rm cum} \approx 5.7\%$ (${\sim}20\%$ per exhaust rest mass; $\eta_{\rm trad}\approx 1.1\%$); continuous thrust achieves $\eta_{\rm cum} \approx 3.7\%$ for the representative matched escape trajectory ($\sim$65\% of the impulse value). All efficiencies are conditional on extraction-with-escape events.

\begin{acknowledgments}
This work is financially supported by VinUniversity under the Environmental Intelligence (CEI) Grant (No. VUNI.CEI.FS\_0009).
\end{acknowledgments}

\appendix

\section{Numerical Validation}
\label{app:diagnostics}

We verified numerical accuracy through multiple diagnostics. We maintain the mass-shell constraint $\mathcal{C} = g^{\mu\nu}p_\mu p_\nu + m^2$ to $|\mathcal{C}| < 10^{-9}$ for single-impulse trajectories and $< 10^{-6}$ for continuous thrust (with projection restoring $< 10^{-14}$). We verify future-directedness ($u^t > 0$) and timelike normalization throughout all trajectories. Energy conservation via 4-momentum balance yields residuals $< 10^{-9}$.

We cross-validated a representative subset of continuous-thrust trajectories with RK45, finding consistent escape/capture outcomes and $\Delta E$ within statistical uncertainty. Boyer--Lindquist coordinates become numerically stiff near $r_+$; integration failures (11.2\% of samples) occur exclusively for trajectories approaching the horizon ($r < 1.1\,r_+$), and we conservatively treat such failures as captures. We find that treating failures as captures yields consistent escape probabilities (Table~\ref{tab:conservative-rates}). Because these failures cluster near the horizon rather than distributing uniformly, the conservative counting may slightly bias estimated threshold locations (e.g.,\ the velocity-onset value), not only lower overall success rates. Future work could validate near-horizon outcomes in a horizon-regular coordinate system (e.g., Kerr--Schild) to further distinguish coordinate-induced integration failures from physical captures.

\begin{table}[htpb]
\centering
\footnotesize
\caption{Simulation accounting breakdown. Phases~1--4 total 104{,}000 trajectories; Phase~5 adds 140{,}000 LHS trajectories for spin-threshold characterization; Phases~6--7 add 76{,}000 additional sweep trajectories. Grand total: 320{,}000. Parameter sweeps use DOP853 (rtol$=10^{-9}$, atol$=10^{-11}$); continuous thrust uses classical RK4 with 4-momentum projection.}
\begin{tabular}{clr}
\hline
Phase & Description & Trajectories \\
\hline
1 & Broad grid ($80\!\times\!80$, 5 spins) & 32{,}000 \\
2 & Focused grid ($60\!\times\!60$, 5 spins) & 18{,}000 \\
3 & Thrust sensitivity ($4\!\times\!4$ grid) & 24{,}000 \\
4 & M.C.\ validation ($3\!\times\!2\!\times\!5{,}000$) & 30{,}000 \\
\hline
5 & Spin threshold ($14\!\times\!10{,}000$ LHS) & 140{,}000 \\
6 & Velocity sweep, Fig.~\ref{fig:thrust-sensitivity} & 40{,}000 \\
7 & Ultra-rel.\ sweep, Fig.~\ref{fig:ultrarel-saturation} & 36{,}000 \\
\hline
& \textbf{Total} & \textbf{320{,}000} \\
\hline
\end{tabular}
\label{tab:simulation-accounting}
\end{table}

\begin{table}[htpb]
\centering
\caption{Conservative extraction-with-escape success rates illustrating the impact of integration failures for representative high-spin cases in a focused sampling domain ($E_0 \in [1.1, 1.3]$, $L_z \in [2.5, 3.5]$) under the periapsis-impulse protocol with $v_e = 0.95c$ and $\delta m_{\rm nom} = 0.20$. P1:``Success (excl.\ int.\ failures)'' excludes integration failures from the denominator; P2: ``Success (failures as captures)'' counts all failures as non-successes. The differences correspond to the observed $\sim$11.2\% failure fraction and support conservatively classifying integration failures as captures near the horizon.}
\begin{tabular}{lccc}
\hline
$a/M$ & P1 & P2 & Diff. \\
\hline
0.90 & 0.38\% & 0.33\% & $-0.05\%$ \\
0.95 & 8.34\% & 7.41\% & $-0.93\%$ \\
0.99 & 10.84\% & 9.63\% & $-1.21\%$ \\
\hline
\end{tabular}
\label{tab:conservative-rates}
\end{table}

\section{Exploratory $a/M=0.99$ Results}
\label{app:exploratory-99}

Table~\ref{tab:thrust-params-99} shows thrust parameter sensitivity results for $a/M = 0.99$, now updated with independent-seed verification ($N = 1{,}000$ per configuration, distinct random seed per $(v_e, \delta m_{\rm nom})$ pair). These results are exploratory and not part of the main evidentiary spine.

\begin{table}[htpb]
\centering
\caption{\textbf{Exploratory:} Thrust parameter sensitivity for $a/M = 0.99$ in the sweet-spot region centered at ($E_0 = 1.22$, $L_z = 3.05$), with $N = 1{,}000$ initial conditions per configuration and distinct random seeds per $(v_e, \delta m_{\rm nom})$ pair (same Gaussian sampling protocol, $\delta m_{\rm nom}$ convention, and $\delta m_{\rm actual}$ discrepancies as Table~\ref{tab:thrust-params}). The original reused-ensemble scan yielded uniformly $86.3\%$ at $v_e = 0.98c$ across all $\delta m_{\rm nom}$; independent-seed verification reveals ${\sim}65$--$70\%$ with mild $\delta m_{\rm nom}$ dependence (4--6 percentage-point spread), confirming that the reused-ensemble rates were upward-biased by a particularly favorable seed. Peak success of $\mathbf{70.0\%}$ at $v_e = 0.98c$, $\delta m_{\rm nom} = 0.4$.}
\begin{tabular}{lcccc}
\hline
$v_e/c$ & $\delta m_{\rm nom} = 0.1$ & $\delta m_{\rm nom} = 0.2$ & $\delta m_{\rm nom} = 0.3$ & $\delta m_{\rm nom} = 0.4$ \\
\hline
0.80 & 0.0\% & 0.0\% & 0.0\% & 0.0\% \\
0.90 & 0.0\% & 0.1\% & 0.1\% & 0.0\% \\
0.95 & 33.3\% & 32.7\% & 31.7\% & 36.2\% \\
0.98 & 68.3\% & 63.7\% & 65.1\% & \textbf{70.0\%} \\
\hline
\end{tabular}
\label{tab:thrust-params-99}
\end{table}

Under the fixed $a/M = 0.95$-centered baseline prior, the relative ranking of $a/M = 0.99$ and $a/M = 0.95$ is not monotone across the parameter space. At $v_e = 0.90c$, $a/M = 0.99$ achieves ${\sim}0\%$ success compared to ${\sim}1\%$ for $a/M = 0.95$; at $v_e = 0.95c$, the ordering depends on $\delta m_{\rm nom}$ (Tables~\ref{tab:thrust-params}--\ref{tab:thrust-params-99}). At $v_e = 0.98c$, $a/M = 0.99$ achieves ${\sim}65$--$70\%$ compared to $69.6\%$ for $a/M = 0.95$, indicating comparable performance under the $a/M = 0.95$-centered prior. This non-monotone behavior is a \emph{fixed-prior artifact}: the sweet-spot center $(E_0,L_z)=(1.22,3.05)$ was derived at $a/M=0.95$ and need not be optimal for $a/M=0.99$, whose escape corridors occupy a different region of phase space. A spin-optimized prior centered on the $a/M = 0.99$ sweet spot would likely shift the balance.

Independent-seed verification confirms mild $\delta m_{\rm nom}$ dependence (4--6 percentage-point spread at $v_e = 0.95c$ and $v_e = 0.98c$), in contrast to the reused-ensemble artifact of identical rates across all $\delta m_{\rm nom}$ columns.

\section*{Data Availability}

Simulation code is available at \url{https://github.com/anindex/penrose_process}. Animated trajectory visualizations are provided in the codebase.

\bibliography{references}

\end{document}